\documentclass[a4paper,UKenglish]{lipics-v2018}

\usepackage{microtype}
\usepackage{amsmath}
\usepackage{amsfonts}
\usepackage{amssymb} 
\usepackage{amscd}
\usepackage{mathrsfs}
\usepackage{mathtools}
\usepackage{amsthm}
\usepackage{color}
\usepackage{tikz}
\usepackage{listings}
\usepackage{pgfplots}
\usepackage{booktabs}
\usetikzlibrary{arrows,automata,positioning,fit,calc,shapes}
\usepackage{xcolor}
\usepackage{pgf}
\usepackage{graphicx}
\usepackage{empheq}
\usepackage{caption}
\usepackage[babel]{csquotes}
\theoremstyle{plain}
\newtheorem{proposition}[theorem]{Proposition}
\newtheorem{conjecture}[theorem]{Conjecture}
\newtheorem{procedure}[theorem]{Procedure}
\title{On randomized generation of slowly synchronizing automata}

\titlerunning{On randomized generation of slowly synchronizing automata}

\author{Costanza Catalano}{Gran Sasso Science Institute\\{Viale Francesco Crispi 7, L'Aquila, Italy}}{costanza.catalano@gssi.it}{}{}

\author{Rapha\"{e}l M. Jungers}{ICTEAM Institute, UCLouvain\\{Avenue Georges Lema\^{i}tres 4-6, Louvain-la-Neuve, Belgium}}{raphael.jungers@uclouvain.be}{}{R. M. Jungers is a FNRS Research Associate. He is supported by the French Community of Belgium, the Walloon Region and the Innoviris Foundation.}

\authorrunning{C. Catalano and R.M. Jungers}

\Copyright{Costanza Catalano and Rapha\"{e}l M. Jungers}

\subjclass{, .}


\category{}

\relatedversion{}

\supplement{}

\funding{}

\acknowledgements{}

\EventEditors{Igor Potapov, Paul Spirakis, and James Worrell}
\EventNoEds{3}
\EventLongTitle{43rd International Symposium on Mathematical Foundations of Computer Science (MFCS 2018)}
\EventShortTitle{MFCS 2018}
\EventAcronym{MFCS}
\EventYear{2018}
\EventDate{August 27--31, 2018}
\EventLocation{Liverpool, GB}
\EventLogo{}
\SeriesVolume{117}
\ArticleNo{48} 

\nolinenumbers 
\hideLIPIcs  

\begin{document}

\maketitle

\begin{abstract}
Motivated by the randomized generation of slowly synchronizing automata, we study automata made of permutation letters and a merging letter of rank $ n\!-	\!1 $. We present a constructive randomized procedure to generate synchronizing automata of that kind with (potentially) large alphabet size based on recent results on \textit{primitive} sets of matrices. We report numerical results showing that our algorithm finds automata with much larger reset threshold than a mere uniform random generation and we present new families of automata with reset threshold of $ \Omega(n^2/4) $. We finally report theoretical results on randomized generation of primitive sets of matrices: a set of permutation matrices with a $ 0 $ entry changed into a $ 1 $ is primitive and has exponent of $ O(n\log n) $ with high probability in case of uniform random distribution and the same holds for a random set of binary matrices 
where each entry is set, independently, equal to $ 1 $ with probability $ p $ and equal to $ 0 $ with probability $ 1-p $, when $ np-\log n\rightarrow\infty $ as $ n\rightarrow\infty $.
 \end{abstract}

\section{Introduction}
A (complete deterministic finite) \emph{automaton} $ \mathcal{A} $ on $ n $ states can be defined as a set of $ m $ binary row-stochastic\footnote{A \textit{binary} matrix is a matrix with entries in $ \lbrace 0,1\rbrace $. A \textit{row-stochastic} matrix is a matrix with nonnegative entries where the entries of each row sum up to $ 1 $. Therefore a matrix is \textit{binary} and \textit{row-stochastic} if each row has exactly one $ 1 $.} matrices $ \lbrace A_1, \dots ,A_m\rbrace $ that are called the \textit{letters} of the automaton. We say that $ \mathcal{A} $ is \textit{synchronizing} if there exists a product of its letters, with repetitions allowed, that has an all-ones column\footnote{A column whose entries are all equal to $ 1 $.} and the length of the shortest of these products is called the \textit{reset threshold} ($rt(\mathcal{A})$) of the automaton.
In other words, an automaton is synchronizing if there exists a word that brings the automaton into a particular state, regardless of the initial one. 
Synchronizing automata appear in different research fields; for example they are often used as models of error-resistant systems \cite{Epp,Chen} and in symbolic dynamics \cite{mateescu}. For a brief account on synchronizing automata and their other applications we refer the reader to \cite{Volk}.
The importance of synchronizing automata also arises from one of the most longstanding open problems in this field, the \v{C}ern\'{y} conjecture, which affirms that any synchronizing automaton on $ n $ states has reset threshold at most $ (n-1)^2 $. If it is true, the bound is sharp due to the existence of a family of $ 2 $-letter automata attaining this value, family discovered by \v{C}ern\'{y} in \cite{Cerny}. 
  Despite great effort, the best upper bound for the reset threshold known so far is 
  $(15617 n^3 + 7500 n^2 + 9375 n - 31250)/93750$, recently obtained by Szyku{\l}a in \cite{Szykula} and thereby beating the 30 years-standing upper bound of $ (n^3-n)/6 $ found by Pin and Frankl in \cite{Frankl,Pin}. 
Better upper bounds have been obtained for certain families of automata and the search for automata attaining quadratic reset threshold within these families have been the subject of several contributions in recent years. These results are (partly) summarized in Table \ref{table}. \\Exhaustive search confirmed the conjecture for small values of $ n $ (see \cite{SlowAutom, BondtDon}).
\small
\begin{table}
\begin{tabular}{lll}
\toprule
\textbf{Classes} & \textbf{Upper b. on $ \mathbf{rt} $}  & \textbf{Families with quadratic $ \mathbf{rt} $} \\
\midrule
Eulerian automata  & $ n^2-3n+3 $ & $ (n^2-3)/2 $ \\
 & Kari \cite{Kari} &  Szyku{\l}a and Vorel \cite{Szykula2016} ($ 4 $ letters)\\
\midrule
Automata with full  & $ 2n^2-6n+5 $ & $ n(n-1)/2 $\\
transition
monoid &Gonze et.\ al.\ \cite{Babai} & Gonze et.\ al.\ \cite{Babai} ($ n $ letters)\\
\midrule
One cluster automata & $n^2-7n+7$ &  $ (n-1)^2\,$ \\
  & B\'{e}al et al. \cite{Beal} & \v{C}ern\'{y} \cite{Cerny} ($ 2 $ letters) \\
\midrule
Strongly connected weakly & $ \lfloor n(n+1)/6\rfloor $ & ?\\
monotone automata    & Volkov \cite{Volkov2007}  &   \\
\midrule
Automata with  & $2(n-1)^2$ &   $ (n-1)^2\,$ \v{C}ern\'{y} \cite{Cerny} ($ 2 $ letters)\\
simple idempotents  & Rystov \cite{Rystsov} &   $ \geq (n^2+3n-6)/4 \quad$ for $\,\, n=4k+3$ \\
& & [Conjectured $ (n^2-1)/2 $]\\
&   &$\geq (n^2+3n-8)/4 \quad$ for $\,\, n=4k+1$  \\
& & [Conjectured $ (n^2-1)/2 $]\\
& &$\geq (n^2+2n-4)/4 \quad$ for $\,\, n=4k$  \\
& & [Conjectured $ (n^2-2)/2 $]\\
& &$\geq (n^2+2n-12)/4\quad $ for $\,\, n=4k+2$ \\
& &[Conjectured $ (n^2-10)/2 $]\\
 &  & \textbf{Our contribution} ($ 3 $ letters)  \\
\bottomrule
\end{tabular}
\caption{Table on upper bounds on the reset threshold for some classes of automata and examples of automata with large reset threshold belonging to these classes, up to date.}
\label{table}
\end{table}
\normalsize
The hunt for a possible counterexample to the conjecture turned out not to be an easy task as well; the search space is wide and calculating the reset threshold is computationally hard 
(see \cite{Epp,Olsch}).
Automata with reset thresholds close to $ (n-1)^2 $, called \textit{extremal} or \emph{slowly synchronizing} automata, are also hard to detect
and not so many families are known; Bondt et.\ al.\ \cite{BondtDon} make a thorough analysis of automata with small number of states and we recall, among others, the families found by Ananichev et al.\ \cite{SlowAutom}, by Gusev and Pribavkina \cite{GusevPriba}, by Kisielewicz and Szyku{\l}a \cite{Szykula2015} and by Dzyga et.\ al.\ \cite{GusevSzikulaDzyga}. These last two examples are, in particular, some of the few examples of slowly synchronizing automata with more than two letters that can be found in the literature. 
Almost all the families of slowly synchronizing automata listed above are closely related to the \v{C}ern\'{y} automaton $ \mathcal{C}(n)=\lbrace a,b\rbrace $, where $ a $ is the cycle over $ n $ vertices and $ b $ the letter that fixes all the vertices but one, which is mapped to the same vertex as done by $ a $; indeed all these families present a letter that is a cycle over $ n $ vertices and the other letters have an action similar to the one of letter $ b $. As these examples seem to have a quite regular structure, it is natural to wonder whether a randomized procedure to generate automata could obtain less structured automata with possibly larger reset thresholds. This probabilistic approach can be rooted back to the work of Erd\H{o}s in the 60's, where he developed the so-called \emph{Probabilistic Method}, a tool that permits to prove the existence of a structure with certain desired properties by defining a suitable probabilistic space in which to embed the problem; for an account on the probabilistic method we refer the reader to \cite{AlonSpencer}. \\The simplest way to randomly generate an automaton of $ m $ letters is to uniformly and independently sample $ m $ binary row-stochastic matrices: unfortunately, Berlinkov first proved in \cite{Berl} that two uniformly sampled random binary row-stochastic matrices synchronize with high probability (i.e.\ the probability that they form a synchronizing automaton tends to $ 1 $ as the matrix dimension tends to infinity), then Nicaud showed in \cite{Nicaud} that they also have reset threshold of order $ O(n\log^3 n) $ with high probability. We say that an automaton is \textit{minimally synchronizing} if any proper subset of its letters is not synchronizing; what just presented before implies that a uniformly sampled random automaton of $ m $ letters has low reset threshold and is \emph{not} minimally synchronizing with high probability. 
Summarizing:
\begin{itemize}
\item slowly synchronizing automata cannot be generated by a mere uniform randomized procedure;
\item minimally synchronizing automata with more than $ 2 $ letters are especially of interest as they are hard to find 
and they do not appear often in the literature, so the behaviour of their reset threshold is still unclear.
\end{itemize}
With this motivation in place, our paper tackles the following questions:

\begin{enumerate}
\item [\textbf{Q1}] Is there a way to randomly generate (minimally) slowly synchronizing automata (with more than two letters)?
\item [\textbf{Q2}] Can we find some automata families with more than two letters, quadratic reset threshold and that do not resemble the \v{C}ern\'{y} family?
\end{enumerate}

 
\textbf{Our Contribution}. 
 In this paper we give positive answers to both questions \textbf{Q1} and \textbf{Q2}. For the first one, we rely on the concept of \textit{primitive} set of matrices, introduced by Protasov and Voynov in \cite{ProtVoyn}: 
 a finite set of matrices with nonnegative entries is said to be \textit{primitive} if there exists a product of these matrices, with repetitions allowed, with all positive entries. A product of this kind is called \emph{positive} and the length of the shortest positive product of a primitive set $ \mathcal{M} $ is called the \textit{exponent} ($ exp(\mathcal{M})$) of the set. Although the Protasov-Voynov primitivity has gained a lot of attention in different fields as in stochastic switching systems \cite{ProtJung} and consensus for discrete-time multi-agent systems \cite{Pierre}, we are interested in its connection with automata theory. In the following, we say that a matrix is \emph{NZ} if it has neither zero-rows nor zero-columns\footnote{Thus a NZ-matrix must have a positive entry in every row and in every column.}; a matrix set is said to be \emph{NZ} if all its matrices are NZ.
\begin{definition}\label{def:assoc_autom}
 Let $ \mathcal{M}\!=\!\lbrace M_1,\dots ,M_m\rbrace $ be a binary NZ-matrix set. 
The \emph{automaton associated to} the set $ \mathcal{M} $ is the automaton $ \mathcal{A}(\mathcal{M}) $ whose letters are all the binary row-stochastic matrices that are entrywise not greater than at least one matrix in $\mathcal{M}$.
\end{definition}

\begin{example}\label{ex}
We here provide an example of a primitive set $ \mathcal{M}=\lbrace M_1,M_2\rbrace $ and the associated automata $ \mathcal{A}(\mathcal{M}) $ and $ \mathcal{A}(\mathcal{M}^T) $ in both their matrix and graph representations (Figure \ref{fig:twoautom}), where $\mathcal{M}^T=\lbrace M^T_1,M^T_2\rbrace  $.\\
$ \mathcal{M}\!=\!\left\lbrace  
\left( \begin{smallmatrix} 0 & 1&0 \\ 1&0&0 \\  0&0&1 \end{smallmatrix}\right) , \left( \begin{smallmatrix} 1 & 0&1 \\  0&0&1 \\ 0&1 & 0 \end{smallmatrix}\right) \right\rbrace $, $ \mathcal{A}(\mathcal{M})\!=\!\left\lbrace  
\left( \begin{smallmatrix} 0 & 1&0 \\ 1&0&0 \\  0&0&1 \end{smallmatrix}\right) , \left( \begin{smallmatrix} 1 & 0&0 \\  0&0&1 \\ 0&1 & 0 \end{smallmatrix}\right) ,\left(  \begin{smallmatrix} 0 & 0&1 \\  0&0&1 \\ 0&1 & 0 \end{smallmatrix}\right) \right\rbrace \!=\! \lbrace a,b,c\rbrace $,\\ $ \mathcal{A}(\mathcal{M}^T)\!=\!\left\lbrace  
\left( \begin{smallmatrix} 0 & 1&0 \\ 1&0&0 \\  0&0&1 \end{smallmatrix}\right) ,\left(  \begin{smallmatrix} 1 & 0&0 \\  0&0&1 \\ 0&1 & 0 \end{smallmatrix}\right) ,\left(  \begin{smallmatrix} 1 & 0&0 \\  0&0&1 \\ 1&0 & 0 \end{smallmatrix}\right) \right\rbrace \!=\! \lbrace a,b,c'\rbrace $.
\begin{figure}[h!]
$ \mathcal{A}(\mathcal{M})\,\, $
\begin{tikzpicture}[shorten >=1pt,node distance=2.5cm,on grid,auto,scale=0.7,transform shape,inner sep=0pt,bend angle=15,line width=0.2mm]


\node[state]    (q_1) {1 };
 			\node[state]    (q_2) [ below right=of q_1] {2};
 			\node[state]          (q_3) [below left=of q_1] {3};

 			\path[->] (q_1) edge	[loop left]	node  {b$ \,$} (q_2)
 			(q_1) edge	[bend right]	 node  {a} (q_2)
 			(q_2) edge [bend right]	node  {a} (q_1)
 			(q_3) edge [loop left]		node  {a$ \,$ } ()
 			(q_2) edge [bend right]		node  {b,c} (q_3)
 			(q_3) edge [bend right]		node  {b,c} (q_2)
 			(q_1) edge 		node  {c} (q_3)
 			;  
\end{tikzpicture}$ \qquad\qquad\qquad\mathcal{A}(\mathcal{M}^T)\,\, $
\begin{tikzpicture}[shorten >=1pt,node distance=2.5cm,on grid,auto,scale=0.7,transform shape,inner sep=0pt,bend angle=15,line width=0.2mm]


\node[state]    (q_1) {1 };
 			\node[state]    (q_2) [ below right=of q_1] {2};
 			\node[state]          (q_3) [below left=of q_1] {3};

 			\path[->] (q_1) edge	[loop left]	node  {b,c'} (q_2)
 			(q_1) edge	[bend right]	 node  {a} (q_2)
 			(q_2) edge [bend right]	node  {a} (q_1)
 			(q_3) edge [loop left]		node  {a$ \,$} ()
 			(q_2) edge [bend right]		node  {b,c'} (q_3)
 			(q_3) edge [bend right]		node  {b} (q_2)
 			(q_3) edge 		node  {c'} (q_1)
 			;  
\end{tikzpicture}
\caption{The automata $ \mathcal{A}(\mathcal{M}) $ and $ \mathcal{A}(\mathcal{M}^T) $ of Example \ref{ex}.}\label{fig:twoautom}
\end{figure}
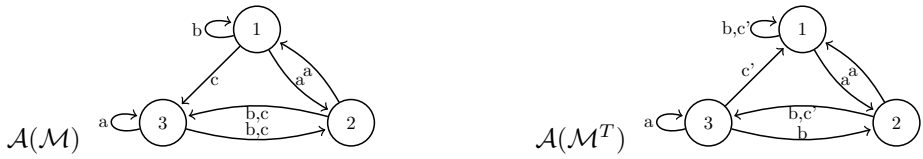
\end{example}
The following theorem summarizes two results proved by Blondel et.\ al.\  (\cite{BlonJung}, Theorems 16-17) and a result proved by Gerencs\'{e}r et al.\ (\cite{GerenGusJung}, Theorem 8). Note that we state it for sets of \emph{binary} NZ-matrices but it more generally holds for any set of NZ-matrices with nonnegative entries; this relies on the fact that in the notion of primitivity what counts is the position of the nonnegative entries within the matrices of the set and not their the actual values. In this case we should add to Definition \ref{def:assoc_autom} the request of setting to $ 1 $ all the positive entries of the matrices of $ \mathcal{M} $ before building $ \mathcal{A}(\mathcal{M}) $. 
\begin{theorem}\label{thm:autom_matrix}
Let $ \mathcal{M}\!=\!\lbrace M_1,\dots ,M_m\rbrace $ a set of binary NZ-matrices of size $ n\times n $ and $ \mathcal{M}^T=\lbrace M^T_1,\dots ,M^T_m\rbrace  $. It holds that $ \mathcal{M} $ is primitive if and only if $ \mathcal{A}(\mathcal{M})$ (equiv. $ \mathcal{A}(\mathcal{M}^T)$) is synchronizing. If $\mathcal{M}  $ is primitive, then it also holds that:
\begin{equation}\label{eq:thmauotm_mat}
\max \Bigl\lbrace rt\bigl(\mathcal{A}(\mathcal{M})\bigr),rt\bigl(\mathcal{A}(\mathcal{M}^T)\bigr)\Bigr\rbrace\leq exp(\mathcal{M}) \leq rt\bigl(\mathcal{A}(\mathcal{M})\bigr)+rt\bigl(\mathcal{A}(\mathcal{M}^T)\bigr)+n-1.
\end{equation}
\end{theorem}

\begin{example}
For the matrix set $\mathcal{M} $ defined in Example \ref{ex}, it holds that $ exp(\mathcal{M})=8 $, $ rt\bigl(\mathcal{A}(\mathcal{M})\bigr)=4 $ and $ rt\bigl(\mathcal{A}(\mathcal{M}^T)\bigr)=2 $.
\end{example}
Theorem \ref{thm:autom_matrix} will be extensively used throughout the paper. 
It shows that primitive sets can be used for generating synchronizing automata and Equation (\ref{eq:thmauotm_mat}) tells us that the presence of a primitive set with large exponent implies the existence of an automaton with large reset threshold; in particular the discovery of a primitive set $ \mathcal{M} $ with $ exp(\mathcal{M})\geq 2(n-1)^2-n+1 $ would disprove the \v{C}ern\'{y} conjecture.
On the other hand, the upper bounds on the automata reset threshold mentioned before imply that $ exp(\mathcal{M})\!=\!O(n^3) $.
\\One advantage of using primitive sets is the Protasov-Voynov characterization theorem (see Theorem \ref{thmProt} in Section \ref{sec:def}) that describes a combinatorial property that a NZ-matrix set must have in order \textit{not} to be primitive: by constructing a primitive set such that each of its proper subsets has this property, we can make it \emph{minimally primitive}\footnote{Thus a \emph{minimally primitive} set is a primitive set that does not contain any proper primitive subset.}.
\\We decided to focus our attention on what we call \textit{perturbed permutation} sets, i.e.\ sets made of permutation matrices (binary matrices having exactly one $ 1 $ in every row and in every column) where a $ 0 $-entry of one of these matrices is changed into a $ 1 $. These sets have many interesting properties: 
\begin{itemize}
\item they have the least number of positive entries that a NZ-primitive set can have, which intuitively should lead to sets with large exponent;
\item the associated automaton $ \mathcal{A}(\mathcal{M}) $ of a perturbed permutation set $ \mathcal{M} $ can be easily computed. It is made of permutation letters and a letter of rank $ n\!-\! 1 $ and its alphabet size is just one unit more than the cardinality of $ \mathcal{M} $;
\item if the matrix set $ \mathcal{M} $ is minimally primitive, the automaton $ \mathcal{A}(\mathcal{M}) $ is minimally synchronizing (or it can be made minimally synchronizing by removing one known letter, as shown in Proposition \ref{prop:minsetsautom});
\item primitivity is easily checked by the Protasov-Voynov algorithm (\cite{ProtVoyn}, Proposition 2), and primitivity of $ \mathcal{M} $ assures that $ \mathcal{A}(\mathcal{M}) $ is synchronizing (Theorem \ref{thm:autom_matrix}).
\end{itemize}
The characterization theorem for primitive sets and the above properties are the main ingredients of our randomized algorithm that finds minimally synchronizing automata of $ 3$ and $4 $ letters (and can eventually be generalized to $ m $ letters); to the best of our knowledge, this is the first time where a constructive procedure for finding minimally synchronizing automata is presented. This is described in Section \ref{sec:minsets} where numerical results are reported. 
The random construction let us also find new families of $ 3 $-letters automata, presented in Section \ref{sec:famaut}, with reset threshold $ \Omega(n^2/4) $ and that do not resemble the \v{C}ern\'{y} automaton, thus answering question \textbf{Q2}.
Finally, in Section \ref{sec:whp} we extend a result on perturbed permutation sets obtained by Gonze et al.\ in \cite{Gonze}: we show that a random perturbed permutation set is primitive with high probability for any matrix size $ n $ (and not just when $ n $ is a prime number as in \cite{Gonze}) and that its exponent is of order $ O(n\log n) $ still with high probability. A further generalization is then presented for sets of random binary matrices: if each entry of each matrix is set to $ 1 $ with probability $ p $ and to $ 0 $ with probability $ 1-p $, independently from each other, then the set is primitive and has exponent of order $ O(n\log n) $ with high probability for any $ p $ such that $ np-\log n\rightarrow \infty $ as $ n\rightarrow \infty $. \\The proofs of the results presented in this paper have been omitted due to length restrictions.

 

\section{Definitions and notation}\label{sec:def}
In this section we briefly go through some definitions and results that will be needed in the rest of the paper.\\
We indicate with $ [n] $ the set $ \lbrace 1,\dots ,n\rbrace $ and with $ S_k $ the set of permutations over $ k$ elements; with a slight abuse of notation $ S_k $ will also denote the set of the $ k\times k $ permutation matrices.\\
An $ n $-state automaton $ \mathcal{A}\!=\!\lbrace A_1,\dots ,A_m\rbrace $ can be represented by a labelled digraph  on $ n $ vertices with a directed edge from vertex $ i $ to vertex $ j $ labelled by $ A_k $ if $ A_k(i,j)=1 $; in this case we also use the notation $ iA_k=j $. 
We remind that a matrix $ M $ is \emph{irreducible} if there does not exist a permutation matrix $ P $ such  that $ PMP^T $ is block-triangular; a set $ \lbrace M_1,\dots ,M_m\rbrace $ is said to be \emph{irreducible} iff the matrix $ \sum_{i=1}^m M_i $ is irreducible. The \textit{directed graph associated to} an $ n\!\times \! n $ matrix $ M $ is the digraph $ D_M $ on $ n $ vertices with a directed edge from $ i $ to $ j $ if $ M(i,j)>0 $. A matrix $ M $ is irreducible if and only if $ D_M $ is strongly connected, i.e.\ if and only if there exists a directed path between any two given vertices.
A \textit{primitive} set $ \mathcal{M} $ is a set of $ m $ matrices $ \lbrace M_1, \dots ,M_m\rbrace $ with nonnegative entries where there exists a product $ M_{i_1}\cdots M_{i_l}>0 $ entrywise, for $ i_1,\dots ,i_l\in [m] $. The length of the shortest of these products is called the \textit{exponent} ($exp(\mathcal{M})$) of the set. Irreducibility is a necessary (but not sufficient) condition for a matrix set to be primitive (see \cite{ProtVoyn}, Section 1). 
Primitive sets of NZ-matrices can be characterized as follows:
\begin{definition}\label{def2}
Let $ \Omega=\dot{\bigcup}^k_{l=1} \Omega_l $ be a partition of $ [n] $ with $ k\geq 2 $. We say that an $ n\times n $ matrix $ M $ has a \textit{block-permutation structure on the partition} $ \Omega $ if there exists a permutation $ \sigma\!\in\!S_k $ such that $\forall \,l\!=\!1,\dots ,k  $ and $\forall\, i\!\in\! \Omega_l $, if $M(i,j)>0  $ then $j\in \Omega_{\sigma(l)}$. We say that a set of matrices \textit{has a block-permutation structure} if there exists a partition on which \textit{all} the matrices of the set have a block-permutation structure.
\end{definition}
\begin{theorem}[\cite{ProtVoyn}, Theorem 1]\label{thmProt}
An irreducible set of NZ matrices of size $ n\times n $ is \emph{not} primitive if and only if the set has a block-permutation structure.
\end{theorem}

We end this section with the last definition and our first observation (Proposition \ref{cor:gonze}).
 \begin{definition}
A matrix $ A $ \textit{dominates} a matrix $ B $ if $ A(i,j)\geq B(i,j),\,\,\forall \, i,j $.
\end{definition}




\begin{proposition}\label{cor:gonze}
Consider an irreducible set $ \lbrace M_1,\dots ,M_m\rbrace $ in which every matrix dominates a permutation matrix. If the set has a block-permutation structure, then all the blocks of the partition must have the same size.
\end{proposition}



\section{Minimally primitive sets and minimally synchronizing automata}\label{sec:minsets}
In this section we focus on \textit{perturbed permutation} sets, i.e.\ matrix sets made of permutation matrices where a $ 0 $-entry of one matrix is changed into a $ 1 $. We represent a set of this kind as $ \mathcal{M}=\lbrace P_1,\dots  ,P_{m-1}, P_m+\mathbb{I}_{i,j}\rbrace $, where $ P_1,\dots ,P_{m} $ are permutation matrices, $ \mathbb{I}_{i,j} $ is a matrix whose $ (i,j) $-th entry is equal to $ 1 $ and all the others entries are equal to $ 0 $ and $ P_m(i,j')=1 $ for a $ j'\neq j $.
The first result states that we can easily generate minimally synchronizing automata 
starting from minimally primitive perturbed permutation sets:


 \begin{proposition}\label{prop:minsetsautom}
Let $ \mathcal{M}=\lbrace P_1, \dots, P_{m-1}, P_m + \mathbb{I}_{i,j}\rbrace $ be a minimally primitive perturbed permutation set and let $j'\neq j$ be the integer such that $ P_m(i,j')=1 $. The automaton $ \mathcal{A}(\mathcal{M}) $ (see Definition \ref{def:assoc_autom}) can be written as $\mathcal{A}(\mathcal{M})=\lbrace P_1, \dots, P_{m-1}, P_m,M\rbrace$ with $ M= P_m + \mathbb{I}_{i,j}- \mathbb{I}_{i,j'} $. If $\mathcal{A}(\mathcal{M})  $ is not minimally sychronizing, then $ \bar{\mathcal{A}}=\lbrace P_1, \dots, P_{m-1}, M\rbrace $ is. 
\end{proposition}


\subsection{A randomized algorithm for constructing minimally primitive sets}\label{sec:algorithm}
If we want to find minimally synchronizing automata, Proposition \ref{prop:minsetsautom} tells us that we just need to generate minimally primitive perturbed permutation sets; in this section we implement a randomized procedure to build them.\\ 
Theorem \ref{thmProt} says that a matrix set is \textit{not} primitive if all the matrices share the same block-permutation structure, therefore a set of $ m $ matrices is minimally primitive iff every subset of cardinality $ m-1 $ has a block-permutation structure on a certain partition; this is the condition we will enforce. 
As we are dealing with perturbed permutation sets, Proposition \ref{cor:gonze} tells us that we just need to consider partitions with blocks of the same size; if the blocks of the partition have size $ n/q $, we call it a \textit{$ q $-partition} and we say that the set has a \textit{$ q $-permutation structure}. Given $ R,C\subset [n] $ and a matrix $ M $, we indicate with $ M(R,C) $ the submatrix of $ M $ with rows indexed by $ R $ and columns indexed by $ C $. The algorithm first generates a set of permutation matrices satisfying the requested block-permutation structures and then a $ 0 $-entry of one of the obtained matrices is changed into a $ 1 $; while doing this last step, we will make sure that such change will preserve all the block-permutation structures of the matrix. We underline that our algorithm finds perturbed permutation sets that, \emph{if} are primitive, are minimally primitive. Indeed, the construction itself only ensures minimality and not primitivity: this latter property has to be verified at the end.
\subsubsection{The algorithm}
For generating a set of $ m $ matrices $ \mathcal{M}\!=\!\lbrace M_1,\dots ,M_m\rbrace $ we choose $ m $ prime numbers $ q_1\geq \dots \geq q_m\geq 2 $ and we set $ n\!=\!\prod_{i=1}^m q_i $. For $ j\!=\!1,\dots ,m $, we require the set $ \lbrace M_1,\dots ,M_{j-1}, M_{j+1},\dots ,M_m\rbrace  $ (the set obtained from $ \mathcal{M} $ by erasing matrix $ M_j $) to have a $ q_j $-permutation structure; this construction will ensure the minimality of the set. More in detail, for all $ j\!=\!1,\dots ,m $ we will enforce the existence of a $ q_j $-partition  $ \Omega_{q_j}\!=\dot{\bigcup}_{i=1}^{q_j}\Omega^j_i $ of $ [n] $ on which, for all $ k\neq j $, the matrix $ M_k $ has to have a block-permutation structure.
As stated by Definition \ref{def2}, this request means that for every $ k=1,\dots ,m $ and for every $ j\neq k $ there must exist a permutation $ \sigma^k_j \in S_{q_j}$ such that for all $ i\!=\!1,\dots ,q_j $ and $ l\neq \sigma^k_j(i) $, $ M_k(\Omega_i^j,\Omega_l^j) $ is a zero matrix.
\\The main idea of the algorithm is to initialize every entry of each matrix to $ 1 $ and then, step by step, to set to $ 0 $ the entries that are not compatible with the conditions that we are requiring. 
As our final goal is to have a set of permutation matrices, at every step we need to make sure that each matrix dominates at least one permutation matrix, despite the increasing number of zeros among its entries. 

\begin{definition}
Given a matrix $ M$ and a $ q$-partition $ \Omega_{q}\!=\dot{\bigcup}_{i=1}^{q}\Omega^q_i $, we say that a permutation $ \sigma \in S_{q}$ is \textit{compatible} with $ M $ and $ \Omega_{q}$ if
for all $i=1,\dots ,q $, there exists a permutation matrix $ Q_i $ such that
\begin{equation}\label{eq:compatible}
M\bigl(\Omega^q_i, \Omega^q_{\sigma(i)}\bigr)\geq Q_i  .
\end{equation} 
\end{definition}

The algorithm itself is formally presented in Listing 1; we here describe in words how it operates. 
Each entry of each matrix is initialized to $ 1 $. The algorithm has two for-loops: the outer one on $ j=1,\dots ,m $, where a $ q_j $-partition $ \Omega_{q_j}=\dot{\bigcup}_{i=1}^{q_j}\Omega^j_i $ of $ [n] $ is uniformly randomly sampled and then the inner one on $k=1,\dots,m$ with $ k\neq j $ where we verify whether there exists a permutation $ \sigma^k_j \in S_{q_j}$ that is compatible with $ M_k $ and $ \Omega_{q_j}$. If it does exist, then we choose one among all the compatible permutations and the algorithm moves to the next step $ k+1 $. 
If such permutation does not exist, then the algorithm exits the inner for-loop and it
randomly selects another $ q_j $-partition $ \Omega'_{q_j} $ of $ [n] $; it then repeats the inner loop for $k=1,\dots,m$ with $ k\neq j $ with this new partition. If after $ T1 $ steps it is choosing a different $ q_j $-partition $ \Omega'_{q_j} $ the existence, for each $ k\neq j $, of a permutation 
$\sigma'^k_j \in S_{q_j} $ that is compatible with $ M_k $ and $\Omega'_{q_j}  $ is not established, 
we stop the algorithm and we say that \emph{it did not converge}. If the inner for-loop is completed, then for each $ k\neq j $ the algorithm modifies the matrix $ M_k $ by keeping unchanged each block $ M_k\bigl(\Omega^j_i, \Omega^j_{\sigma_j^k(i)}\bigr) $ for $ i=1,\dots ,q_j $ and by setting to zero all the other entries of $ M_k $, where $ \sigma_j^k $ is the selected compatible permutation; the matrix $ M_k $ has now a block-permutation structure over the sampled partition $\Omega_{q_j}  $. The algorithm then moves to the next step $ j+1 $. If it manages to finish the outer for-loop, we have a set of binary matrices with the desired block-permutation structures. We then just need to select a permutation matrix $ P_k\leq M_k $ for every $ k=1,\dots ,m $ and then to randomly change a $ 0 $-entry into a $ 1 $ in one of the matrices without modifying its block-permutation structures: this is always possible as the blocks of the partitions are non trivial and a permutation matrix has just $ n  $ positive entries. We finally check whether the set is primitive.\\
Here below we present the procedures that the algorithm uses:

\begin{alphaenumerate}

\item $ [p,P]=Extractperm(M,met) $\\
This is the key function of the algorithm. It returns $p\!=\!1$ if the matrix $M$ dominates a permutation matrix, it returns $p\!=\!0$ and $ P\!=\!M $ otherwise. In the former case it also returns a permutation matrix $P$ selected among the ones dominated by $M$ according to $met$; if $met=2$ the matrix $P$ is sampled uniformly at random, while if $met=3$ we make the choice of $P$ deterministic. More in detail,
the procedure works as follows: we first count the numbers of $ 1 $s in each column and in each row of the matrix $M$. We then consider the row or the column with the least number of $ 1 $s; if this number is zero we stop the procedure and we set $p=0$, as in this case $M$ does not dominate a permutation matrix. If this number is strictly greater than zero, we choose one of the $ 1 $-entries of the row or the column attaining this minimum: if $met=2$ (\textbf{method 2}) the entry is chosen uniformly at random while if $met=3$ (\textbf{method 3}) we take the first $ 1 $-entry in the lexicographic order. Suppose that such $ 1 $-entry is in position $ (i,j) $: we set to zero all the other entries in row $ i $ and column $ j $ and we iterate the procedure on the submatrix obtained from $M$ by erasing row $ i $ and column $ j $.
We can prove that this procedure is well-defined and in at most $ n $ steps it produces the desired output: $p=0$ if and only if $M$ does not dominate a permutation matrix and, in case $p=1$, method 2 indeed sample uniformly one of the permutations dominated by $M$, while method 3 is deterministic and the permutation obtained usually has its $ 1 $s distributed around the main diagonal. 
Method 3 will play an important role in our numerical experiments in Section \ref{sec:numexp} and in the discovery of new families of automata with quadratic reset threshold in Section \ref{sec:famaut}.

\item $[a,A]=DomPerm(M,\Omega,met)$\\
It returns $a\!=\!1$ if there exists a permutation compatible with the matrix $M$ and the partition $\Omega=\dot{\bigcup}_{i=1}^{q}\Omega^q_i$, it returns $a\!=\!0$ and $A\!=\!M$ otherwise. In the former case it chooses one of the compatible permutations, say $ \sigma $, according to $met$ and returns the matrix $A$ equal to $ M $ but the entries not in the blocks defined 
by (\ref{eq:compatible}) that are set to zero; $A$ has then a block-permutation structure on $\Omega$. 
More precisely, $DomPerm$ acts in two steps: it first defines a $ q\times q $ matrix $ B $ such that, for all $ i,k=1,\dots ,q $,\\
$\qquad \qquad \qquad
B(i,k)=\begin{cases}
1 & \text{if } M(\Omega_i^q,\Omega_k^q) \text{ dominates a permutation matrix }\\
0 & \text{ otherwise}
\end{cases}
;$\\this can be done by calling $ExtractPerm$ with input $ M(\Omega_i^q,\Omega_k^q) $ and $met$ for all $ i,k=1,\dots ,q $. At this point, asking if there exists a permutation compatible with $M$ and $\Omega$ is equivalent of asking if $ B $ dominates a permutation matrix. Therefore, the second step is to call again $[p,P]=ExtractPerm(B,met)$: if $p=0$ we set $a=0$ and $A=M$, while if $p=1$ we set $a=1$ and $A$ as described before with $\sigma=P$ (i.e.\ $ \sigma(i)=j $ iff $ P(i,j)=1 $); indeed the permutation $P$ is one of the permutations compatible with $M$ and $\Omega$.

\item $M\!set=Addone(P_1,...,P_m)$\\
It changes a $ 0$-entry of one of the matrices $ P_1,...,P_m $ into a $ 1 $ preserving all its block-permutation structures. The matrix and the entry are chosen uniformly at random and the procedure iterates the choice till it finds a compatible entry (which always exists); it then returns the final perturbed permutation set $M\!set$.
\item $pr=Primitive(M\!set)$\\
It returns $pr\!=\!1$ if the matrix set $M\!set\!=\! \lbrace M_1,\dots ,M_m\rbrace $ is primitive and $pr=0$ otherwise. It first verifies if the set is irreducible by checking the strong connectivity of the digraph $D_N$ where $N\!=\!\sum_{i=1}^k M_i$ (see Section \ref{sec:def}) via breadth-first search on every node, 
then if the set is irreducible, primitivity is checked by the Protasov-Voynov algorithm (\cite{ProtVoyn}, Section 4).
\end{alphaenumerate}
All the above routines have polynomial time complexity in $ n $, apart from routine $Primitive$ that has time complexity $ O(mn^2) $. 

\begin{remark}\label{rem:alpin}
\begin{enumerate}
\item In all our numerical experiments the algorithm always converged, i.e.\ it always ended before reaching the stopping value $ T1 $, for $ T1 $ large enough. This is probably due to the fact that the matrix dimension $ n $ grows exponentially as the number of matrices $ m $ increases, which produces enough degrees of freedom. We leave the proof of this fact for future work. 

\item A recent work of Alpin and Alpina \cite{Alpin} generalizes Theorem \ref{thmProt} for the characterization of primitive sets 
to sets that are allowed to be reducible and the matrices to have zero columns but not zero rows. Clearly, automata fall within this category.
Without going into many details (for which we refer the reader to \cite{Alpin}, Theorem 3), Alpin and Alpina show that an $ n $-state automaton is \textit{not} synchronizing if and only if there exist a partition $ \dot{\bigcup}_{j=1}^s\Omega_j $ of $ [n] $ such that it has a block-permutation structure on a \textit{subset} of that partition. This characterization is clearly less restrictive: 
 it just suffices to find a subset $ I\subset [s] $ such that for each letter $ A $ of the automaton there exists a permutation $ \sigma\in S_I $ such that for all $ i\in I $, if $ A(i,j)=1 $ then $ j\in \Omega_{\sigma(i)} $. 
 Our algorithm could leverage this recent result in order to directly construct minimal synchronizing automata.  We also leave this for future work.
\end{enumerate}
\end{remark}

\begin{lstlisting}[caption={Algorithm for finding minimally primitive sets.},label=list:8-6,captionpos=t,abovecaptionskip=-\medskipamount]
Input: q_1,...,q_m,T1,met
Initialize M_1,...,M_m as all-ones matrices
for j:=1 to m do 
	t1=0
	while t1<T1 do	
		t1=t1+1
		choose a q_j-partition Omega_j
		for k=1 to m and k!=j do
		   [a,A_k]=DomPerm(M_k,Omega_j,met)
		   if a==0
			 exit inner for-loop 		
		   end
		end
		if a==1
		   exit while-loop
		end	
	end
	if t1==T1
	    display 'does not converge', exit procedure	
	else
       	    set M_k=A_k for all k=1,...,m and k!=j
	end
end
for i:=1 to m do
    [p_i,P_i]=Extractperm(M_i,met)
end
Mset=Addone(P_1,...,P_m)
pr=Primitive(Mset)
return Mset, pr
\end{lstlisting}


 

\subsection{Numerical results}\label{sec:numexp}
We have seen that once we have a minimally primitive perturbed permutation set, it is easy to generate a minimally synchronizing automaton from it, as stated by Proposition \ref{prp:auotmashape}. 
Our goal is to generate automata with large research threshold, but checking this property on many randomly generated instances is prohibitive. Indeed, we recall that computing the reset threshold of an automaton is in general NP-hard \cite{Epp}.
Instead, as a proxy for the reset threshold, we compute the \emph{diameter of the square graph}, which we now introduce:
\begin{definition}\label{def:sg}
The \emph{square graph} $ S(\mathcal{A}) $ of an $ n $-state automaton $ \mathcal{A} $ is the labelled directed graph with vertex set $ V\!=\!\lbrace (i,j): 1\leq i\leq j\leq n\rbrace $ and edge set $ E $ such that $ e\!=\!\lbrace (i,j), (i',j')\rbrace\!\in \! E $ if there exists a letter $ A\!\in\!\mathcal{A} $ such that $ A(i,i')>0 $ and $ A(j,j')>0 $, or $ A(i,j')>0 $ and $ A(j,i')>0 $. In this case, we label the directed edge $ e $ by $ A $ (multiple labels are allowed). A vertex of type $ (i,i) $ is called a \emph{singleton}.
\end{definition}
A well-known result (\cite{Volk}, Proposition 1) states that an automaton is synchronizing if and only if in its square graph there exists a path from any non-singleton vertex to a singleton one; the proof of this fact also implies that 
\begin{equation}\label{eq:sg}
diam\bigl(S(\mathcal{A})\bigr)\leq rt(\mathcal{A})\leq n\cdot diam\bigl(S(\mathcal{A})\bigr),
\end{equation}
where $ diam\bigl(S(\mathcal{A})\bigr) $ denotes the \emph{diameter} of $S(\mathcal{A})  $ i.e. the maximum length of the shortest path between any two given vertices, taken over all the pairs of vertices. 
The diameter can be computed in polynomial time, namely $ O(mn^2) $ with $ m $ the number of letters of the automaton. 

\begin{figure}
\includegraphics[scale=0.3]{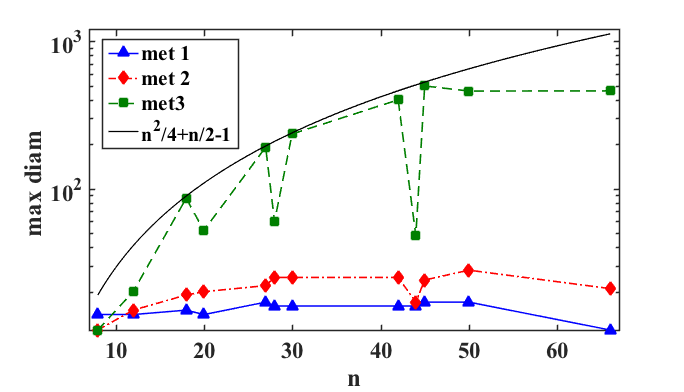}
\includegraphics[scale=0.3]{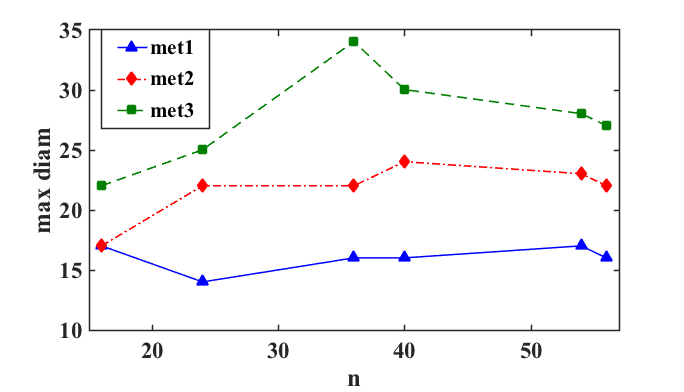}
\caption{Comparison of our methods (met 2 and 3) with the naive method (met 1) with respect to the maximal diameter found on $ 50n^2 $ iterations.  Left: $n$ is the product of three prime numbers; the $ y $ axis is in logarithmic scale. Right: $n$ is the product of four prime numbers.}
\label{fig:3mat}
\end{figure}

We now report our numerical results based on the diameter of the square graph. We compare three methods of generating automata: we call \textbf{method 1} the uniform random generation of 2-letter automata made of one permutation matrix and a matrix of rank $ n-1 $, while \textbf{method 2} and \textbf{method 3}, already introduced in the previous paragraph, refer to the different ways of extracting a permutation matrix from a binary one in our randomized construction. 
We set $ T1\!=\!1000 $ and for each method and each choice of $ n $ we run the algorithm $it(n)= 50n^2 $ times, thus producing each time $ 50n^2 $ sets. This choice for $ it(n) $ has been made by taking into account two facts: on one hand, it is desirable to keep constant the rate $ it(n)/k_m(n) $ between the number of sampled sets $ it(n) $ and the cardinality $ k_m(n) $ of the set of the perturbed permutation sets made of $ m $ matrices. Unfortunately, $ k_m(n+1)/k_m(n) $ grows approximately as $ n^m $ and so $ k_m(n) $ explodes very fast. On the other hand, we have to deal with the limited computational speed of our computers. The choice of $it(n)= 50n^2 $ comes as a compromise between these two issues, at least when $ n\leq 70 $.\\
Among the $ it(n) $ generated sets, we select the primitive ones and we generate their associated automata (Definition \ref{def:assoc_autom}); we then check which ones are not minimally synchronizing and we make them minimally synchronizing by using Proposition \ref{prop:minsetsautom}. Finally, we compute the square graph diameter of all the minimally synchronizing automata obtained.
Figure \ref{fig:3mat} reports on the $ y $ axis the maximal square graph diameter found for each method and for each matrix dimension $ n $ when $ n $ is the product of three prime numbers (left picture) and when it is a product of four prime numbers (right picture). We can see that our randomized construction manages to reach higher values of the square graph diameter than the mere random generation; in particular, method 3 reaches quadratic diameters in case of three matrices. 

\begin{figure}
\includegraphics[scale=0.29]{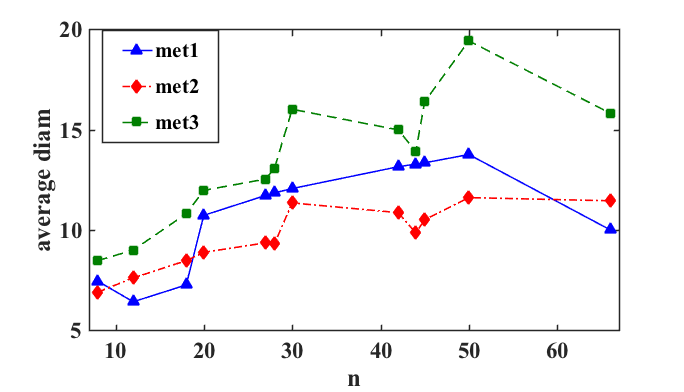}
\includegraphics[scale=0.31]{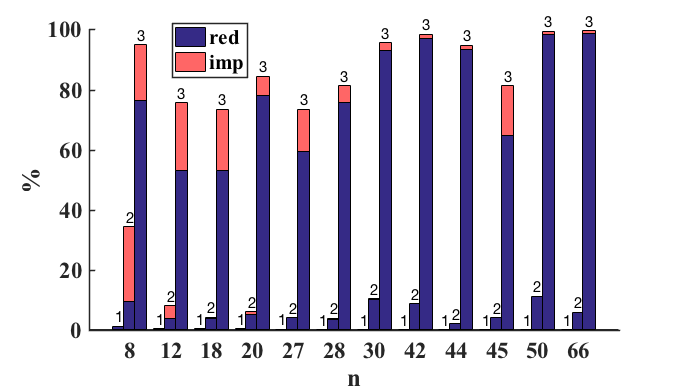}
\caption{Left: Average diameter found by methods 1, 2 and 3 when $n$ is the product of three prime numbers. Right: Percentage of nonprimitive sets (divided into reducible and imprimitive sets) generated by methods 1, 2 and 3 (indicated above each bar) when $n$ is the product of three prime numbers. For instance, on sets of dimension $ n=20 $, method $1$ generates $ 0.35\% $ of nonprimitive sets ($ 0.35\% $ reducible, $ 0\% $ imprimitive), method $2$ generates $ 6.15\% $ of nonprimitive sets ($ 5.18\% $ reducible, $ 0.97\% $ imprimitive) and method $3$ generates $ 84.5\% $ of nonprimitive sets ($ 77.9\% $ reducible, $ 6.6\% $ imprimitive).}
\label{fig:avgredimp}
\end{figure}

We also report in Figure \ref{fig:avgredimp} (left) the behavior of the \emph{average} diameter of the minimally synchronizing automata generated on $ 50n^2 $ iterations when $ n $ is the product of three prime numbers: we can see that in this case method $ 2 $ does not perform better than method $1$, while method $ 3 $ performs just slightly better. This behavior could have been expected since our primary goal was to randomly generate \emph{at least one} slowly synchronizing automata; this is indeed what happens with method $3$, that manages to reach quadratic reset thresholds most of the times.\\
A remark can be done on the percentage of the generated sets that are \emph{not} primitive; this is reported in Figure \ref{fig:avgredimp} (right), where we divide nonprimitive sets into two categories: reducible sets and \emph{imprimitive} sets, i.e.\ irreducible sets that are not primitive. We can see that the percentage of nonprimitive sets generated by method 1 goes to $ 0 $ as $ n $ increases, behavior that we expected (see Section \ref{sec:whp}, Theorem \ref{Thm:primperm}), while method 2 seems to always produce a not negligible percentage of nonprimitive sets, although quite small. The behavior is reversed for method $ 3 $: most of the generated sets are not primitive. This can be interpreted as a good sign. Indeed, nonprimitive sets can be seen as sets with \emph{infinite} exponent; as we are generating a lot of them with method $3$, we intuitively should expect that, when a primitive set is generated, it has high chances to have large diameter.\\
The slowly synchronizing automata found by our randomized construction are presented in the following section. We believe that some parameters of our construction, as the way a permutation matrix is extracted from a binary one or the way the partitions of $ [n] $ are selected, could be further tuned or changed in order to generate new families of slowly synchronizing automata; for example, we could think about selecting the ones in the procedure $Extractperm$ according to a given distribution. We leave this for future work.

\section{New families of automata with quadratic reset threshold}\label{sec:famaut}
We present here four new families of (minimally synchronizing) $ 3 $-letter automata with square graph diameter of order $ \Omega(n^2/4)$, which represents a lower bound for their reset threshold. These families are all made of two symmetric permutation matrices and a matrix of rank $ n-1 $ that merges two states and fixes all the others (a perturbed identity matrix): they thus lie within the class of automata \textit{with simple idempotents}, class introduced by Rystsov in \cite{Rystsov} in which every letter $ A $ of the automaton is requested either to be a permutation or to satisfy $ A^2=A $. These families have been found via the randomized algorithm described in Section \ref{sec:algorithm} using the deterministic procedure to extract a permutation matrix from a binary one (method 3). 
The following proposition shows that primitive sets made of a perturbed identity matrix and two symmetric permutations must have a very specific shape; we then present our families, prove closed formulas for their square graph diameter and finally state a conjecture on their reset thresholds. With a slight abuse of notation we identify a permutation matrix $ Q $ with its underlying permutation, that is we say that $ Q(i)=j $ if and only if $ Q(i,j)=1 $; the identity matrix is denoted by $ I $. Note that a permutation matrix is symmetric if and only if its cycle decomposition is made of fixed points and cycles of length $ 2 $. 
\begin{proposition}\label{prp:auotmashape}
Let $ \mathcal{M}_{ij}=\lbrace \bar{I}_{ij}, Q_1, Q_2\rbrace $ be a matrix set of $ n\times n $ matrices where $\bar{I}_{ij}=I+\mathbb{I}_{ij}$, $ j\neq i $, is a perturbed identity and $ Q_1 $ and $ Q_2 $ are two symmetric permutations. If $ \mathcal{M} $ is irreducible then, up to a relabelling of the vertices, $ Q_1 $ and $ Q_2 $ have the following form:\\
- if $ n $ is even
\small
\begin{equation}\label{saus1}
Q_1(i)=\begin{cases}
1 &\text{if } i=1\\
i+1 &\text{if } i \text{ even, }2\leq i\leq n-2\\
i-1 &\text{if } i \text{ odd, }3\leq i\leq n-1\\
n &\text{if } i=n\\
\end{cases},\quad
Q_2(i)=\begin{cases}
i-1 &\text{if } i \text{ even}\\
i+1 &\text{if } i \text{ odd}\\
\end{cases}
\end{equation}
\normalsize
or
\small
\begin{equation}\label{saus2}
Q_1(i)=\begin{cases}
n &\text{if } i=1\\
i+1 &\text{if } i \text{ even, }2\leq i\leq n-2\\
i-1 &\text{if } i \text{ odd, }3\leq i\leq n-1\\
1 &\text{if } i=n\\
\end{cases},\quad
Q_2(i)=\begin{cases}
i-1 &\text{if } i \text{ even}\\
i+1 &\text{if } i \text{ odd}\\
\end{cases}
\end{equation}\normalsize
- if $ n $ is odd
\small
\begin{equation}\label{sausodd}
Q_1(i)=\begin{cases}
1 &\text{if } i=1\\
i+1 &\text{if } i \text{ even}\\
i-1 &\text{if } i \text{ odd, }3\leq i\leq n\\
\end{cases},\quad
Q_2(i)=\begin{cases}
i-1 &\text{if } i \text{ even}\\
i+1 &\text{if } i \text{ odd, }1\leq i\leq n\!-\!2\\
n &\text{if } i=n\\
\end{cases}.
\end{equation}
\end{proposition}


\begin{proposition}\label{prop:noprimset}
A matrix set $ \mathcal{M}_{ij}=\lbrace \bar{I}_{ij}, Q_1, Q_2\rbrace $ of type (\ref{saus2}) is never primitive.
\end{proposition}




\begin{definition}\label{defn:Aij}
We define $ \mathcal{A}_{ij}=\lbrace \underline{I}_{ij}, Q_1,Q_2\rbrace $ to be the associated automaton $ \mathcal{A}(\mathcal{M}_{ij}) $ of $ \mathcal{M}_{ij} $ (see Definition \ref{def:assoc_autom}), where $ \underline{I}_{ij}=I+\mathbb{I}_{ij}-\mathbb{I}_{ii} $. 
\end{definition}

It is clear that $ \mathcal{A}_{ij}$ is with simple idempotents. Figure \ref{fig:autom} represents $ \mathcal{A}_{1,6} $ with $ n=8 $. We set now  $ \mathcal{E}_n=\mathcal{A}_{1,n-2} $ for $ n=4k $ and $ k\geq 2 $, $ \mathcal{E}'_n=\mathcal{A}_{1,n-4} $ for $ n\!=\!4k+2 $ and $ k\geq 2 $, $ \mathcal{O}_n=\mathcal{A}_{\frac{n-1}{2},\frac{n+1}{2}} $ for $ n=4k+1 $ and $ k\geq 1 $, $ \mathcal{O}'_n=\mathcal{A}_{\frac{n-1}{2},\frac{n+1}{2}} $ for $ n=4k+3 $ and $ k\geq 1 $. The following theorem holds:

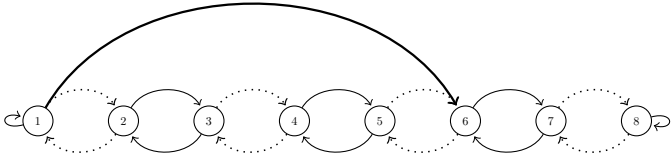
\begin{figure}
\begin{tikzpicture}[shorten >=1pt,node distance=2.5cm,on grid,auto,scale=0.45,transform shape,inner sep=0.8pt,bend angle=60,every state/.style={text=black},every node/.style={text=black}]
  \node[state]  (q_0)                      {1};
  \node[state]          (q_1) [ right=of q_0] {2};
  \node[state]          (q_2) [right =of q_1] {3};
  \node[state](q_3) [ right=of q_2] {4};
  \node[state](q_4) [ right=of q_3] {5};
  \node[state](q_5) [ right=of q_4] {6};
  \node[state](q_6) [ right=of q_5] {7};
  \node[state](q_7) [ right=of q_6] {8};
  \path[->, line width=0.3mm] (q_0) edge [bend left]				node        {} (q_5);        
   \path[->, dotted,line width=0.2mm ](q_0) edge [bend left]				node        {} (q_1)
  (q_2) 	edge [bend left]	           		node  {} (q_3)
  (q_3) edge [bend left]				node        {} (q_2)
  (q_1) 	edge [bend left]	           		node  {} (q_0)
  (q_4) edge [bend left]				node        {} (q_5)
  (q_6) 	edge [bend left]	           		node  {} (q_7)
  (q_7) edge [bend left]				node        {} (q_6)
  (q_5) 	edge [bend left]	           		node  {} (q_4) ;
\path[->] (q_0) edge [loop left]			node  {} ()	
          (q_1) edge [bend left]				node        {} (q_2)
		(q_3) 	edge [bend left]	           		node  {} (q_4)
		(q_4) edge [bend left]				node        {} (q_3) 
		(q_2) 	edge [bend left]	           		node  {} (q_1)		
(q_5) edge [bend left]				node        {} (q_6)
		(q_7) 	edge [loop right]			node  {} () 	 
		(q_6) 	edge [bend left]	           		node  {} (q_5)
		;	
\end{tikzpicture}
\caption{The automata $ \mathcal{A}_{1,6} $ with $ n=8 $; $ rt(\mathcal{A}_{1,6}) $=$31  $. Dashed arrows refer to matrix $ Q_2 $, normal arrows to matrix $ Q_1 $ and bold arrows to matrix $ \underline{I}_{1,6} $ where its selfloops have been omitted.}\label{fig:autom}
\end{figure}

\begin{theorem}\label{thm:autfam}
The automaton $ \mathcal{E}_n$ has square graph diameter (SGD) of $ (n^2+2n-4)/4 $, $ \mathcal{E}'_n$ has SGD of $ (n^2+2n-12)/4 $, $ \mathcal{O}_n$ has SGD of $ (n^2+3n-8)/4 $ and $ \mathcal{O}'_n$ has SGD of $ (n^2+3n-6)/4$. Therefore all the families $ \mathcal{E}_n$, $ \mathcal{E}'_n$, $ \mathcal{O}_n$ and $ \mathcal{O}'_n$ have reset threshold of $ \Omega(n^2/4) $.
\end{theorem}

\begin{figure}[h!]
\begin{tikzpicture}[shorten >=1pt,node distance=1.3cm,on grid,auto,scale=0.45,transform shape,inner sep=0.4pt,bend angle=70,every state/.style={text=black},every node/.style={text=black}]
  \node[state] [red] (q_4)                      {4,5};
  \node[state]          (q_3) [ above=of q_4] {3,6};
  \node[state]          (q_2) [above=of q_3] {2,7};
   \node[state]          (q_1) [above=of q_2] {1,8};

  \node[state](q_5) [ right=of q_1 , xshift=3.6cm] {7,8};
  \node[state](q_6) [ right=of q_2, xshift=3.6cm] {6,8};
  \node[state](q_7) [ right=of q_3, xshift=3.6cm] {5,7};
   \node[state](q_8) [ right=of q_4, xshift=3.6cm] {4,6};
  \node[state](q_9) [ below=of q_8 ] {3,5};
  \node[state](q_10) [ below=of q_9] {2,4};
  \node[state](q_11) [ below=of q_10] {1,3};
  \node[state](q_12) [ below=of q_11] {1,2};
  
  \node[state](q_13) [ right=of q_5 , xshift=3.6cm] {2,3};
  \node[state](q_14) [ right=of q_6, xshift=3.6cm] {1,4};
  \node[state](q_15) [ right=of q_7, xshift=3.6cm] {1,5};
   \node[state](q_16) [ right=of q_8, xshift=3.6cm] {2,6};
   \node[state](q_17) [ right=of q_9 , xshift=3.6cm] {3,7};
  \node[state](q_18) [ right=of q_10, xshift=3.6cm] {4,8};
  \node[state](q_19) [ right=of q_11, xshift=3.6cm] {5,8};
   \node[state](q_20) [ right=of q_12, xshift=3.6cm] {6,7};
   

\node[state](q_21) [ right=of q_13 , xshift=3.6cm] {3,4};
  \node[state](q_22) [ right=of q_14, xshift=3.6cm] {2,5};
  \node[state](q_23) [ right=of q_15, xshift=3.6cm] {1,6};
   \node[state](q_24) [ right=of q_16, xshift=3.6cm] {1,7};
   \node[state](q_25) [ right=of q_17 , xshift=3.6cm] {2,8};
  \node[state](q_26) [ right=of q_18, xshift=3.6cm] {3,8};
  \node[state](q_27) [ right=of q_19, xshift=3.6cm] {4,7};
   \node[state](q_28) [ right=of q_20, xshift=3.6cm] {5,6};

   \node[state](q_29) [ right=of q_21 , xshift=2cm] {6,6};
   
 \path[->, line width=0.3mm] (q_1) edge[red] node {} (q_6)
 (q_11) edge			node        {} (q_3)	
  (q_12)  edge[red]				node        {} (q_16)
  (q_14) edge 			node        {} (q_8)	
(q_15)  edge	[red]		node        {} (q_23)	
  (q_23) edge[red]			node        {} (q_29)
	(q_24) edge				node        {} (q_20)
  ;
  \path[->, dotted,line width=0.2mm ]
  (q_1) edge [bend left]				node        {} (q_2)
  (q_2) edge [bend left,red]				node        {} (q_1)	
  (q_3) edge [bend left]				node        {} (q_4)
  (q_4) edge [bend left,red]				node        {} (q_3)	
  (q_5) edge [loop above ]				node        {} ()
	(q_6) edge [bend left,red]				node        {} (q_7)	
	(q_7) edge [bend left]				node        {} (q_6)
	(q_8) edge [bend left,red]				node        {} (q_9)	
	(q_9)edge [bend left]				node        {} (q_8)	
	(q_10) edge [bend left,red]				node        {} (q_11)
	(q_11) edge [bend left]				node        {} (q_10)	
	(q_12) edge [loop below ]				node        {} ()
	(q_13) edge [bend left]				node        {} (q_14)
	(q_14) edge [bend left]				node        {} (q_13)
	(q_15) edge [bend left]				node        {} (q_16)
	(q_16) edge [bend left,red]				node        {} (q_15)
	(q_17) edge [bend left]				node        {} (q_18)
	(q_18) edge [bend left]				node        {} (q_17)
	(q_19) edge [bend left]				node        {} (q_20)
	(q_20)	edge [bend left]				node        {} (q_19)
	(q_21) edge [loop above ]				node        {} ()
	(q_22) edge [bend left]				node        {} (q_23)
	(q_23) edge [bend left]				node        {} (q_22)
	(q_24) edge [bend left]				node        {} (q_25)
	(q_25) edge [bend left]				node        {} (q_24)	
	(q_26) edge [bend left]				node        {} (q_27)
	(q_27) edge [bend left]				node        {} (q_26)
	(q_28) edge [loop below ]				node        {} ()	
  ;
\path[->] (q_1) edge [loop above ]				node        {} ()						
(q_2) edge [bend left]				node        {} (q_3)	
(q_3) edge [bend left,red]				node        {} (q_2)			
(q_4) edge [loop below ]				node        {} ()		
(q_5) edge [bend left]				node        {} (q_6)
(q_6) edge [bend left]				node        {} (q_5)	
(q_7) edge [bend left,red]				node        {} (q_8)	
(q_8) edge [bend left]				node        {} (q_7)	
(q_9) edge [bend left,red]				node        {} (q_10)	
(q_10) edge [bend left]				node        {} (q_9)	
(q_11) edge [bend left,red]				node        {} (q_12)
(q_12) edge [bend left]				node        {} (q_11)
(q_13) edge [loop above ]				node        {} ()				
(q_14) edge [bend left]				node        {} (q_15)
(q_15) edge [bend left]				node        {} (q_14)	
(q_16) edge [bend left]				node        {} (q_17)
(q_17) edge [bend left]				node        {} (q_16)	
(q_18) edge [bend left]				node        {} (q_19)
(q_19) edge [bend left]				node        {} (q_18)	
(q_20) edge [loop below ]				node        {} ()	
(q_21) edge [bend left]				node        {} (q_22)
(q_22) edge [bend left]				node        {} (q_21)	
(q_23) edge [bend left]				node        {} (q_24)
(q_24) edge [bend left]				node        {} (q_23)	
(q_25) edge [bend left]				node        {} (q_26)	
(q_26) edge [bend left]				node        {} (q_25)		
(q_27) edge [bend left]				node        {} (q_28)	
(q_28) edge [bend left]				node        {} (q_27)	
					;	
\end{tikzpicture}
\caption{Square graph of automaton $ \mathcal{E}_{8} $, $diam\bigl(S(\mathcal{E}_{8})\bigr)=  19$. Normal arrows refer to matrix $ Q_1 $, dotted arrows to matrix $ Q_2 $ and bold arrows to matrix $ \underline{I}_{1,6}$, where its selfloops have been omitted. The red path is the diameter.}\label{figSG8}
\end{figure}
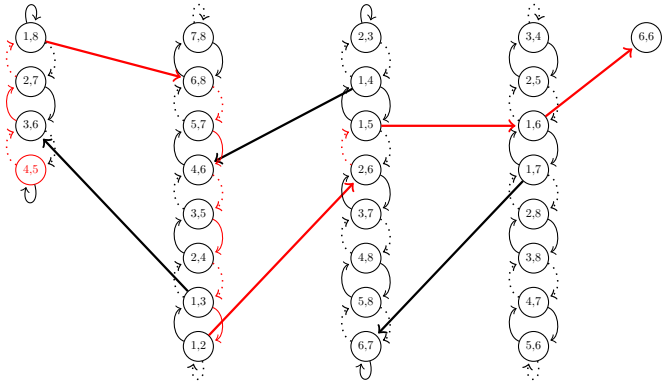

Figure \ref{figSG8} represents the square graph of the automaton $ \mathcal{E}_{8} $, where its diameter is colored in red. All the singletons but the one that belongs to the diameter have been omitted.

\begin{conjecture}
The automaton $ \mathcal{E}_n$ has reset threshold of $ (n^2-2)/2 $, $ \,\mathcal{E}'_n$ has reset threshold of $ (n^2-10)/2 $ and $ \mathcal{O}_n$ and $ \mathcal{O}'_n$ have reset threshold of $ (n^2-1)/2 $. Furthermore, they represent the automata with the largest possible reset threshold among the family $ \mathcal{A}_{ij} $ for respectively $ n=4k $, $ n=4k+2 $, $ n=4k+1 $ and $ n=4k+3 $.
\end{conjecture}

We end this section by remarking that, despite the fact that the randomized construction for minimally primitive sets presented in Section \ref{sec:minsets} works just when the matrix size $ n $ is the product of at least three prime numbers, here we found an extremal automaton of quadratic reset threshold for \emph{any} value of $ n $.


\section{Primitivity with high probability}\label{sec:whp}
We call \textit{random perturbed permutation} set a perturbed permutation set of $ m \!\geq\! 2$ matrices
constructed with the following randomized procedure:
\begin{procedure}\label{proc}
\begin{enumerate}
\item We sample $ m $ permutation matrices $ \lbrace P_1,\dots ,P_m\rbrace $ independently and uniformly at random from the set $ S_n $ of all the permutations over $ n $ elements;
\item A matrix $ P_i $ is uniformly randomly chosen from the set $ \lbrace P_1,\dots ,P_m\rbrace $. Then, one of its $ 0 $-entry is uniformly randomly selected among its $ 0 $-entries and changed into a $ 1 $. It becomes then a perturbed permutation matrix $ \bar{P}_i $;
\item The final random perturbed permutation set is the set $ \lbrace P_1,\dots ,P_{i-1},\bar{P}_i, P_{i+1},\dots ,P_m\rbrace $.
\end{enumerate}
\end{procedure}
This procedure is also equivalent to choosing independently and uniformly at random $ m-1 $ permutation matrices from $ S_n $ and one \textit{perturbed} permutation matrix from $ \bar{S}_n $ with $ \bar{S}_n=\lbrace\bar{P}: \bar{P}\!=\!P+\mathbb{I}_{i,j}, \,P\!\in\! S_n,\, P(i,j')=1,\, j\neq j'\rbrace $. We say that a property $ X $ holds for a random matrix set with \textit{high probability} if the probability that property $ X $ holds tends to $ 1 $ as the matrix dimension $ n $ tends to infinity. 

\begin{theorem}\label{Thm:primperm}
A random perturbed permutation set constructed via Procedure \ref{proc} is primitive and has exponent of order $ O(n\log n) $ with high probability.
\end{theorem}

Theorem \ref{Thm:primperm} can be extended to random sets of binary matrices. 
It is clear that focusing just on binary matrices is not restrictive as in the definition of primitivity what counts is just the position of the positive entries within the matrices and not their actual values.
Let $ B(p) $ denote a random binary matrix where each entry is independently set to $ 1 $ with probability $ p $ and to $ 0 $ with probability $ (1-p) $ and let $ \mathcal{B}_m(p) $ denote a set of $ m\geq 2 $ matrices obtained independently in this way. Under some mild assumptions over $ p $, we still have primitivity with high probability:

\begin{theorem}\label{thm:binarymat}
For any fixed integer $ m\geq 2 $, if $np- \log n\longrightarrow \infty $ as $ n\rightarrow\infty$, then $ \mathcal{B}_m(p) $ is primitive and $exp\bigl(\mathcal{B}_m(p)\bigr)= O(n\log n) $ with high probability.
\end{theorem}

It is interesting to compare this result with the one obtained by Gerencs\'{e}r et al.\ in \cite{GerenGusJung}:  they prove that, if $ exp(n) $ is the maximal value of the exponent among all the binary primitive sets of $ n\times n $ matrices, then
$\lim_{n\rightarrow\infty} (\log exp(n))/ n=(\log 3)/3$.
This implies that, for $ n $ big enough, there must exist some primitive sets whose exponent is close to $ 3^{n/3} $, but these sets must be very few as Theorem \ref{thm:binarymat} states that they are almost impossible to be detected by a mere random generation.
We conclude with a result on when a set of two random binary matrices is \textit{not} primitive with high probability:
\begin{proposition}\label{prop:notprim}
For any fixed integer $ m\geq 2 $, if $ 2np-\log n\longrightarrow -\infty $ as $ n\rightarrow\infty$, then $ \mathcal{B}_m(p) $ is reducible with high probability. This implies that $ \mathcal{B}_m(p) $ is not primitive with high probability.
\end{proposition}


\section{Conclusion}
In this paper we have proposed a randomized construction for generating slowly minimally synchronizing automata. Our strategy relies on a recent characterization of primitive sets (Theorem \ref{thmProt}), together with a construction (Definition \ref{def:assoc_autom} and Theorem \ref{thm:autom_matrix}) allowing to build (slowly minimally) synchronizing automata from (slowly minimally) primitive sets. We have obtained four new families of automata with simple idempotents with reset threshold of order $ \Omega(n^2/4) $. The \textit{primitive sets approach} to synchronizing automata seems promising and we believe that out randomized construction could be further refined and tweaked in order to produce other interesting automata with large reset threshold, for example by changing the way a permutation matrix is extracted by a binary one. 
As mentioned at the end of Section \ref{sec:algorithm}, it would be also of interest to apply the minimal construction directly to automata by leveraging the recent result of Alpin and Alpina (\cite{Alpin}, Theorem 3).

\section*{Acknowledgments} The authors thank Fran\c{c}ois Gonze and Vladimir Gusev 
for significant suggestions and fruitful discussions on the topic.

\bibliography{biblOnprobprim}

\begin{thebibliography}{10}

\bibitem{AlonSpencer}
N.~Alon and J.~H. Spencer.
\newblock {\em The Probabilistic Method}.
\newblock Wiley Publishing, 4th edition, 2016.

\bibitem{Alpin}
Yu. Alpin and V.~Alpina.
\newblock Combinatorial properties of entire semigroups of nonnegative
  matrices.
\newblock {\em Journal of Mathematical Sciences}, 207(5):674--685, 2015.

\bibitem{SlowAutom}
D.~S. Ananichev, M.~V. Volkov, and V.~V. Gusev.
\newblock Primitive digraphs with large exponents and slowly synchronizing
  automata.
\newblock {\em Journal of Mathematical Sciences}, 192(3):263--278, 2013.

\bibitem{Beal}
M.-P. B\'{e}al, M.~V. Berlinkov, and D.~Perrin.
\newblock A quadratic upper bound on the size of a synchronizing word in
  one-cluster automata.
\newblock In {\em International Journal of Foundations of Computer Science},
  pages 277--288, 2011.

\bibitem{Berl}
M.V. Berlinkov.
\newblock {\em On the Probability of Being Synchronizable}, pages 73--84.
\newblock Springer International Publishing, 2016.

\bibitem{BlonJung}
V.D. Blondel, R.M. Jungers, and A.~Olshevsky.
\newblock On primitivity of sets of matrices.
\newblock {\em Automatica}, 61:80--88, 2015.

\bibitem{Chen}
Y.-B. Chen and D.~J. Ierardi.
\newblock The complexity of oblivious plans for orienting and distinguishing
  polygonal parts.
\newblock {\em Algorithmica}, 14(5):367--397, 1995.

\bibitem{Pierre}
P.-Y. Chevalier, J.M. Hendrickx, and R.M. Jungers.
\newblock Reachability of consensus and synchronizing automata.
\newblock In {\em 4th IEEE Conference on Decision and Control}, pages
  4139--4144, 2015.

\bibitem{BondtDon}
M.~de~Bondt, H.~Don, and H.~Zantema.
\newblock Dfas and pfas with long shortest synchronizing word length.
\newblock In {\em Developments in Language Theory}, pages 122--133, 2017.

\bibitem{Epp}
D.~Eppstein.
\newblock Reset sequences for monotonic automata.
\newblock {\em SIAM Journal on Computing}, 19(3):500--510, 1990.

\bibitem{Frankl}
P.~Frankl.
\newblock An extremal problem for two families of sets.
\newblock {\em European Journal of Combinatorics}, (3):125 -- 127, 1982.

\bibitem{Friedman}
J.~Friedman, A.~Joux, Y.~Roichman, J.~Stern, and J.-P. Tillich.
\newblock {\em The action of a few random permutations on r-tuples and an
  application to cryptography}, pages 375--386.
\newblock Springer Berlin Heidelberg, 1996.

\bibitem{GerenGusJung}
B.~Gerencs{\'{e}}r, V.~V. Gusev, and R.~M. Jungers.
\newblock Primitive sets of nonnegative matrices and synchronizing automata.
\newblock {\em Siam Journal on Matrix Analysis and Applications}, 39(1):83--98,
  2018.

\bibitem{Gonze}
F.~Gonze, B.~Gerencs\'{e}r, and R.M. Jungers.
\newblock Synchronization approached through the lenses of primitivity.
\newblock In {\em 35th Benelux Meeting on Systems and Control}.

\bibitem{Babai}
F.~Gonze, V.V. Gusev, B.~Gerencs{\'e}r, R.M. Jungers, and M.V. Volkov.
\newblock {\em On the Interplay Between Babai and {\v{C}}ern{\'y}'s
  Conjectures}, pages 185--197.
\newblock Springer International Publishing, 2017.

\bibitem{Graham}
A.~J. Graham and D.~A. Pike.
\newblock A note on thresholds and connectivity in random directed graphs.
\newblock {\em Atlantic Electronic Journal of Mathematics}, 3(1):1--5, 2008.

\bibitem{GusevPriba}
V.~V. Gusev and E.~V. Pribavkina.
\newblock Reset thresholds of automata with two cycle lengths.
\newblock In {\em Implementation and Application of Automata}, pages 200--210,
  2014.

\bibitem{randgraphs}
S.~Janson, T.~Luczak, and A.~Rucinski.
\newblock {\em Random Graphs}.
\newblock Wiley Series in Discrete Mathematics and Optimization. Wiley, 2011.

\bibitem{Kari}
J.~Kari.
\newblock Synchronizing finite automata on eulerian digraphs.
\newblock {\em Theoretical Computer Science}, 295(1):223 -- 232, 2003.

\bibitem{Szykula2015}
A.~Kisielewicz and M.~Szyku{\l}a.
\newblock Synchronizing automata with extremal properties.
\newblock In {\em Mathematical Foundations of Computer Science 2015}, pages
  331--343, 2015.

\bibitem{mateescu}
A.~Mateescu and A.~Salomaa.
\newblock Many-valued truth functions, \v{C}ern\'{y}’s conjecture and road
  coloring.
\newblock In {\em EATCS Bull.}, page 134–150, 1999.

\bibitem{GusevSzikulaDzyga}
M.~Michalina~Dzyga, R.~Ferens, V.V. Gusev, and M.~Szyku{\l}a.
\newblock Attainable values of reset thresholds.
\newblock In {\em 42nd International Symposium on Mathematical Foundations of
  Computer Science}, volume~83, pages 40:1--40:14, 2017.

\bibitem{Nicaud}
C.~Nicaud.
\newblock Fast synchronization of random automata.
\newblock In {\em Approximation, Randomization, and Combinatorial
  Optimization}, volume~60 of {\em Leibniz International Proceedings in
  Informatics}, pages 43:1--43:12, 2016.

\bibitem{Olsch}
J.~Olschewski and M.~Ummels.
\newblock {\em The Complexity of Finding Reset Words in Finite Automata}, pages
  568--579.
\newblock Springer Berlin Heidelberg, 2010.

\bibitem{Pin}
J.-E. Pin.
\newblock On two combinatorial problems arising from automata theory.
\newblock In {\em Proceedings of the International Colloquium on Graph Theory
  and Combinatorics}, volume~75, pages 535--548, 1983.

\bibitem{ProtJung}
V.Yu. Protasov and R.M. Jungers.
\newblock Lower and upper bounds for the largest lyapunov exponent of matrices.
\newblock {\em Linear Algebra and its Applications}, 438:4448--4468, 2013.

\bibitem{ProtVoyn}
V.Yu. Protasov and A.S. Voynov.
\newblock Sets of nonnegative matrices without positive products.
\newblock {\em Linear Algebra and its Applications}, 437:749--765, 2012.

\bibitem{Rystsov}
I.~K. Rystsov.
\newblock Estimation of the length of reset words for automata with simple
  idempotents.
\newblock {\em Cybernetics and Systems Analysis}, 36(3):339--344, 2000.

\bibitem{Szykula}
M.~Szyku{\l}a.
\newblock Improving the upper bound the length of the shortest reset words.
\newblock In {\em STACS}, 2018.

\bibitem{Szykula2016}
M.~Szyku{\l}a and V.~Vorel.
\newblock An extremal series of eulerian synchronizing automata.
\newblock In {\em Developments in Language Theory}, pages 380--392, 2016.

\bibitem{Cerny}
J.~\v{C}ern\'{y}.
\newblock Pozn\'{a}mka k homog\'{e}nnym eksperimentom s kone\v{c}n\'{y}mi
  automatami.
\newblock {\em Matematicko-fysikalny Casopis SAV}, (14):208 -- 216, 1964.

\bibitem{Volkov2007}
M.~V. Volkov.
\newblock Synchronizing automata preserving a chain of partial orders.
\newblock In {\em Implementation and Application of Automata}, pages 27--37,
  2007.

\bibitem{Volk}
M.V. Volkov.
\newblock {\em Synchronizing automata and the \v{C}ern\'{y} conjecture}, volume
  5196, pages 11--27.
\newblock Springer.

\end{thebibliography}

\appendix
\section{Appendix}

\begin{lstlisting}[caption={Procedure for extracting a permutation matrix from a binary one.},label=list,captionpos=t,abovecaptionskip=-\medskipamount]
Input: M, met
n= size of M
P=M, p=1
I=[1,2,..,n]
J=[1,2,..,n]
for i:=1 to n do
  v1= vector of the number of 1s in the rows of P indexed by I
  v2= vector of the number of 1s in the columns of P indexed by J
  v=[v1,v2]
  sort v in ascending order
  if v(1)==0
    p=0, P=M, exit procedure   
  else
    if v(1) belongs to v1
	 choose a 1-entry in row v(1) according to met
	 j= column index of the 1-entry chosen 
	 set to 0 all the other entries in P in row v(1) and column j
	 delete v(1) from I
	 delete j from J
    else
	 choose a 1-entry in column v(1) according to met
	 j= row index of the 1-entry chosen
         set to 0 all the other entries in P in column v(1) and row j
	 delete v(1) from J
	 delete j from I
    end    
  end
end
return p, P
\end{lstlisting}

\begin{proof} \textbf{of Proposition \ref{cor:gonze}.}\\
Let $ Q_i$ be the permutation matrix dominated by $ M_i $; if $ M_i $ has a block-permutation structure on a given partition, so does $ Q_i $ on the same partition. Theorem 2 in \cite{Gonze} states that if a set of permutation matrices has a block-permutation structure then all the blocks of the partition must have the same size, so the proposition follows. 
\end{proof}

\begin{proof} \textbf{of Proposition \ref{prop:minsetsautom}.}\\
Suppose $ \mathcal{A}(\mathcal{M}) $ is not minimal; the only matrix we can possibly delete from the set without losing the property of being synchronized is $ P_m $. Indeed, we cannot delete $ M $ as all the others are permutation matrices. For $ i=1,\dots ,m-1 $, let $ \mathcal{M}_i$ be the set obtained by erasing $ P_i $ from $ \mathcal{M} $; by hypothesis, $ \mathcal{M}_i $ is not primitive so the automaton $ \mathcal{A}(\mathcal{M}_i)$ is not synchronizing. But $ \mathcal{A}(\mathcal{M}_i)$ is indeed the automaton obtained by erasing $ P_i $ from $ \mathcal{A}(\mathcal{M}) $, so $ \bar{\mathcal{A}} $ has to be synchronizing and minimal.
\end{proof}

\begin{proof} \textbf{of Proposition \ref{prp:auotmashape}.}\\
$ \mathcal{M} $ is irreducible if and only if the digraph $ D $ induced by matrix $ \bar{I}_{ij}+Q_1+Q_2 $ is strongly connected (see Section \ref{sec:def}). For $D$ to be strongly connected, the digraph induced by $ Q_1+Q_2 $ must be strongly connected as $ Q_1 $ and $ Q_2 $ are permutation matrices and the matrix $ \bar{I}_{ij} $ just adds a single edge that is not a selfloop to that digraph.
Let us now consider vertex $ 1 $: there must exist a matrix in the set $ \lbrace Q_1,Q_2\rbrace $ that sends it to another vertex; let it be $ Q_2 $ (without loss of generality) and label this vertex with $ 2 $. As $ Q_2 $ must be symmetric, we have $ Q_2(1)=2 $ and $ Q_2(2)=1 $. This implies that $ Q_1 $ needs to send vertex $ 2 $ to some vertex other than $ 1 $ as otherwise the graph would not be strongly connected; we label this vertex with $ 3 $ and so we have $ Q_1(2)=3 $ and $ Q_1(3)=2 $. By repeating this reasoning, we end up with $ Q_1 $ and $ Q_2 $ of the form (\ref{saus1}) or (\ref{saus2}) if $ n $ is even and of the form (\ref{sausodd}) if $ n $ is odd. 
\end{proof}

\begin{proof} \textbf{of Proposition \ref{prop:noprimset}.}\\
Due to symmetry of digraph $ D_{Q_1+Q_2} $, up to a relabelling of the vertices we can assume without loss of generality that $ i\!=\! 1 $. If $ j $ is odd, all the three matrices have a block-permutation structure over the partition $\left\lbrace \lbrace 1,3,\dots ,n-1\rbrace,\lbrace 2,4,\dots ,n\rbrace \right\rbrace $, while if $ j $ is even they have a block-permutation structure over the partition 
$
\bigl\lbrace \lbrace 1,k\rbrace ,\lbrace 2,k-1\rbrace ,\dots ,\lbrace \frac{k}{2},\frac{k}{2}+1\rbrace, \lbrace k+1,n\rbrace , \lbrace k+2,n-1\rbrace , \dots $  $
\dots ,\lbrace \frac{n+k}{2},\frac{n+k}{2}+1\rbrace\bigr\rbrace .
$
By Theorem \ref{thmProt}, the set cannot be primitive. 
\end{proof}

\begin{figure}
\begin{tikzpicture}[shorten >=1pt,node distance=1.3cm,on grid,auto,scale=0.5,transform shape,inner sep=0.4pt,bend angle=70,every state/.style={text=black},every node/.style={text=black}]

\node at (0,6) {\huge{$ \mathbf{ C_0}$}};
\node at (5,6) {\huge{$\mathbf{ C_1 }$}};
\node at (10,6) {\huge{$ \mathbf{C_{\frac{n-4}{2}}} $}};
\node at (14,6) {\huge{$ \mathbf{\dots} $}};
\node at (19,6) {\huge{$\mathbf{ C_{\frac{n}{2}-1} }$}};

  \node[state] [red]  (q_4)                      {$\frac{n}{2}$, $\frac{n}{2}$+1};
  \node[state]          (q_3) [ above=of q_4] {3,n-2};
  \node[state]          (q_2) [above=of q_3, shape =rectangle] {$ \vdots $};
   \node[state]          (q_1) [above=of q_2] {1,n};

  \node[state](q_5) [ right=of q_1 , xshift=3.6cm] {};
  \node[state](q_6) [ right=of q_2, xshift=3.6cm] {n-2,n};
  \node[state](q_7) [ right=of q_3, xshift=3.6cm] {};
   \node[state](q_8) [ right=of q_4, xshift=3.6cm] {n-4,n-2};
  \node[state](q_9) [ below=of q_8,shape =rectangle ] {$ \vdots $};
  \node[state](q_11) [ below=of q_9] {1,3};
  \node[state](q_12) [ below=of q_11] {1,2};
  
  \node[state](q_13) [ right=of q_5 , xshift=3.6cm] {};
  \node[state](q_14) [ right=of q_6, xshift=3.6cm] {1,n-4};
  \node[state](q_15) [ right=of q_7, xshift=3.6cm] {1,n-3};
   \node[state](q_16) [ right=of q_8, xshift=3.6cm] {2,n-2};
   \node[state](q_17) [ right=of q_9 , xshift=3.6cm,shape =rectangle] {$ \vdots $};
   \node[state](q_20) [ right=of q_11, xshift=3.6cm] {n-2,n-1};
   
 \node[state](q_30) [ right=of q_16, xshift=3cm, shape =rectangle] {$\cdots $};

\node[state](q_21) [ right=of q_13 , xshift=8cm] {};
  \node[state](q_22) [ right=of q_14, xshift=8cm,shape =rectangle] {$\vdots $};
  \node[state](q_23) [ right=of q_15, xshift=8cm] {5,n-2};
   \node[state](q_24) [ right=of q_16, xshift=8cm,shape =rectangle] {$\vdots $};
   \node[state](q_25) [ right=of q_17 , xshift=8cm] {1,n-1 };
  \node[state](q_26) [ below=of q_25] {1,n-2 };
  \node[state](q_27) [ below=of q_26,shape =rectangle] {$\vdots $};
   \node[state](q_28) [ below=of q_27] {$ \frac{n}{2}$-1,$\frac{n}{2} $};

   \node[state](q_29) [ right=of q_21 , xshift=2cm] {n-2,n-2};
   
 \path[->, line width=0.3mm] (q_1) edge[red] node {} (q_6)
 (q_11) edge			node        {} (q_3)	
  (q_12)  edge[red]				node        {} (q_16)
  (q_14) edge 			node        {} (q_8)	
(q_15)  edge	[red]		node        {} (q_30)	
  (q_26) edge	[red]		node        {} (q_29)
  (q_30) edge				node        {} (q_20)
	(q_30) edge	[red]			node        {} (q_26)
	(q_25) edge			node        {} (q_30)
  ;
  \path[->, dotted,line width=0.2mm ]
  (q_1) edge [bend left]				node        {} (q_2)
  (q_2) edge [bend left,red]			node        {} (q_1)	
  (q_3) edge [bend left]				node        {} (q_4)
  (q_4) edge [bend left,red]				node        {} (q_3)	
  (q_5) edge [loop above ]				node        {} ()
	(q_6) edge [bend left,red]				node        {} (q_7)	
	(q_7) edge [bend left]				node        {} (q_6)
	(q_8) edge [bend left,red]			node        {} (q_9)	
	(q_9)edge [bend left]				node        {} (q_8)	
	(q_9) edge [bend left,red]				node        {} (q_11)
	(q_11) edge [bend left]				node        {} (q_9)	
	(q_12) edge [loop below ]				node        {} ()
	(q_13) edge [bend left]				node        {} (q_14)
	(q_14) edge [bend left]				node        {} (q_13)
	(q_15) edge [bend left]				node        {} (q_16)
	(q_16) edge [bend left,red]			node        {} (q_15)
	(q_17) edge [bend left]				node        {} (q_20)
	(q_20)	edge [bend left]				node        {} (q_17)
	(q_21) edge [loop above ]				node        {} ()
	(q_22) edge [bend left]				node        {} (q_23)
	(q_23) edge [bend left]				node        {} (q_22)
	(q_24) edge [bend left]				node        {} (q_25)
	(q_25) edge [bend left]				node        {} (q_24)	
	(q_26) edge [bend left]				node        {} (q_27)
	(q_27) edge [bend left]				node        {} (q_26)
	(q_28) edge [loop below ]				node        {} ()	
  ;
\path[->] (q_1) edge [loop above ]				node        {} ()						
(q_2) edge [bend left]				node        {} (q_3)	
(q_3) edge [bend left,red]				node        {} (q_2)			
(q_4) edge [loop below ]				node        {} ()		
(q_5) edge [bend left]				node        {} (q_6)
(q_6) edge [bend left]				node        {} (q_5)	
(q_7) edge [bend left,red]				node        {} (q_8)	
(q_8) edge [bend left]				node        {} (q_7)	
(q_11) edge [bend left,red]				node        {} (q_12)
(q_12) edge [bend left]				node        {} (q_11)
(q_13) edge [loop above ]				node        {} ()				
(q_14) edge [bend left]				node        {} (q_15)
(q_15) edge [bend left]				node        {} (q_14)	
(q_16) edge [bend left]				node        {} (q_17)
(q_17) edge [bend left]				node        {} (q_16)	
(q_20) edge [loop below ]				node        {} ()	
(q_21) edge [bend left]				node        {} (q_22)
(q_22) edge [bend left]				node        {} (q_21)	
(q_23) edge [bend left]				node        {} (q_24)
(q_24) edge [bend left]				node        {} (q_23)	
(q_25) edge [bend left]				node        {} (q_26)	
(q_26) edge [bend left]				node        {} (q_25)		
(q_27) edge [bend left]				node        {} (q_28)	
(q_28) edge [bend left]				node        {} (q_27)	
					;	
\end{tikzpicture}
\caption{Square graph of the family $ \mathcal{E}_{n} $. There are $ n/2 $ chains, the first one has $ n/2 $ vertices, the others $ n $ vertices; the missing chains and vertices are represented by boxes with dots.  Bold lines refer to matrix $ \underline{I}$, dotted lines to matrix $ Q_2 $ and normal lines to matrix $ Q_1 $. The red path is the diameter. }\label{figSG}
\end{figure}
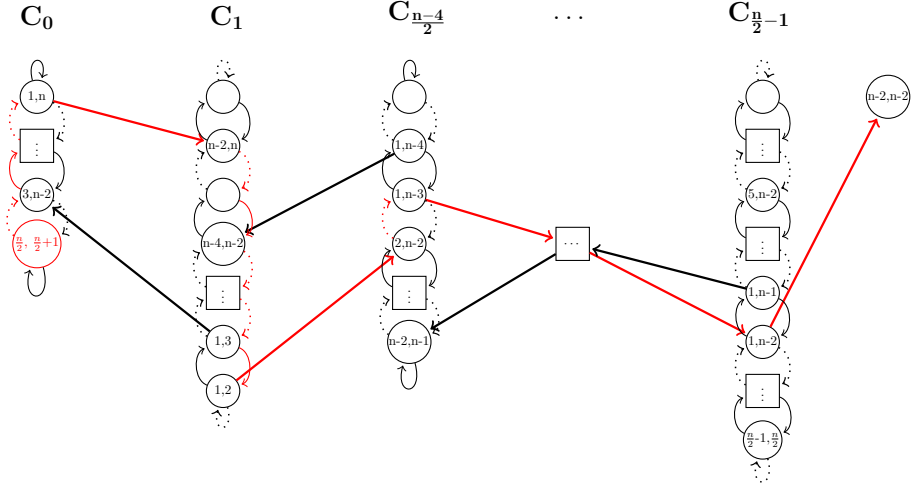

\begin{proof} \textbf{of Theorem \ref{thm:autfam}.}\\
We prove the theorem just for family $ \mathcal{E}_n$; the other square graph diameters can be obtained by similar reasoning. In the following we describe the shape of $ S(\mathcal{A}_{1,n-2} )$ with $ n=4k $ in order to compute its diameter: we invite the reader to refer to Figure \ref{figSG} during the proof. In its description we omit the singleton vertices as the only link in $ S(\mathcal{A}_{1,n-2} )$ between a non-singleton vertex and a singleton one is the edge connecting $ (1,n-2) $ to $ (n-2,n-2) $. 
We set $ \underline{I}=\underline{I}_{1,n-2} $ to ease the notation. If we do not consider the merging letter $ \underline{I} $, the digraph $S(\mathcal{A}_{1,n-2}\setminus \lbrace \bar{I}\rbrace ) $ (singletons omitted) is disconnected and has $ n/2 $ strongly connected components: $ C_0 $ of size $ n/2 $ and $ C_1,\dots C_{n/2-1} $ of size $ n$. The component $ C_0 $ is made of the vertices $ \lbrace (1+s, n-s): s=0,\dots ,n/2 -1\rbrace $ while component $ C_i $ is made of the vertices $ \lbrace (i,i+1),(i-1,i+2),\dots ,(1,2i),(1,2i+1),(2,2i+2),\dots ,(n-i,n-i+1)\rbrace $ for $ 1\leq i \leq n/2-1 $: these components look like ``chains'' due to the symmetry of $ Q_1 $ and $ Q_2 $ (see Figure \ref{figSG}). Component $ C_ 0$ contains the vertices $ (1,n) $ and $ (3,n-2) $, while component $ C_i $ contains vertices $ (1,2i),(1,2i+1) $ for $ 1\leq i\leq n/2\!-\!1 $. Furthermore, $ C_1 $ contains the vertices $ (n-4,n-2)$ and $(n-2,n) $, $ C_{n/2-1} $ contains the vertices $ (1,n-2)$ and $(4,n-2) $ and $ C_i $ contains the vertices $ (n-2i-2,n-2)$ and $(n-2i+3,n-2) $ for $ 2\leq i\leq n/2\!-\!2 $. The matrix $ \underline{I}$ connects the components $\lbrace C_i\rbrace$ by linking vertex $ (1,a) $ to vertex $ (a,n-2) $ for every $a=2,\dots ,n$. 
Figure \ref{fig:diagram} shows how the components $ \lbrace C_i\rbrace $ are linked together for $  2\leq i \leq n/2-1 $: an arrow between two vertices means that there exist a word mapping the first vertex to the other, a number next to the arrow represents the length of such word if the two vertices belong to the same component while arrows connecting vertices from different components are labelled by $ \underline{I} $. Bold vertices represent the ones that are linked by $ \underline{I} $ to other chains. How $ C_0 $ is connected to $ C_1 $ is directly shown in Figure \ref{figSG}.
The digraph $ S(\mathcal{A}_{1,n-2}) $ is thus formed by ``layers'' represented by the components $\lbrace C_i\rbrace $ where $ C_0 $ is the farthest from the singleton $ (n-2, n-2) $ and $ C_{n/2-1}$ is the closest, as shown in Figure \ref{figSG}. In order to compute its diameter we need to measure the length of the shortest path from vertex $ (n/2,n/2+1) $ to vertex $ (n-2,n-2) $, which is colored in red in Figure \ref{figSG}. This means that for $ 1\leq i\leq n/2 $ we have to compute the distance $ d_i $  between vertices $ (2i,n-2) $ and $ (1,n-2i-1) $ or between vertices $ (2i+1,n-2) $ and $ (1,n-2i+2) $, depending which one is part of that path.
In view of the diagram in Figure \ref{fig:diagram}, as the sequence of components from the farthest to the closest to the singleton $ (n-2,n-2) $ is $ C_0,C_1,C_{\frac{n-4}{2}},C_3,C_{\frac{n-8}{2}}, C_5,C_{\frac{n-12}{2}},\dots  $, we have the following sequence for the $ d_i $s:
\[
d_0\!=\!\frac{n}{2}-1,\,d_1\!=\! n-2,\,d_{\frac{n-4}{2}}\!=\!1,\,d_3\!=\! n-3, \,d_{\frac{n-8}{2}}\!=\!5,\,d_5\!=\!n-7,\,d_{\frac{n-12}{2}}\!=\!9, \dots
\]
Observe that $ d_{2i}+d_{2i+1}= n-2 $ for $ 1\leq i\leq n/4 -1 $. Since the number of edges labelled by $ \underline{I} $ that appear in the path is $ n/2 $, finally the diameter is equal to
\[
diam(S(\mathcal{A}_{1,n-2} ))=\frac{n}{2}+ \sum_{k=0}^{\frac{n}{2}} d_k= \frac{n^2}{4}+\frac{n}{2}-1 .
\]
 \end{proof}

\begin{figure}
\begin{align*}
\small
\begin{CD}
\mathbf{C_{\frac{(n-2i+2)}{2}}} :      @.    \mathbf{C_i}: @. \mathbf{C_{\frac{(n-2i-2)}{2}}}: \\
      @.     @. \\
      @.   (1,2i)       @>\underline{I}>> (2i,n-2)\\
      @.     @AA{1}A       @V{2i-1}VV \\
      (2i+1,n-2)  @<\underline{I}<<   (1,2i+1) @.     \mathbf{(1,n-2i-1)} \\
    @VVV       @AA{n-2i-3}A       @V{1}VV \\
   @V{2i}VV        (n-2i-2,n-2)       @<\underline{I}<< (1,n-2i-2) \\
   @VVV			@AA{5}A				@.\\
  (1,n-2i+3)  @>\underline{I}>> (n-2i+3,n-2) @.\\
   @V{1}VV  @. @.\\
  \mathbf{ (1,n-2i+2)}@. @.\\
\end{CD}
\end{align*}
\caption{Diagram on how the components $ \lbrace C_i\rbrace $ in the proof of Theorem \ref{thm:autfam}  are linked together.}\label{fig:diagram}
\end{figure}

We here state a group-theoretic result of Friedman et al.\ \cite{Friedman} that we will need for the proof of Theorem \ref{Thm:primperm}.

\begin{theorem}[Friedman et al.\ \cite{Friedman}]\label{thm:perm}
For any $ r\geq 1 $ and $ d\geq 2 $ and for a uniform random sample of $d$ permutation matrices $ P_1,\dots ,P_d $ from $ S_n $, the following property $ F_r $ happens with high probability: for any two $r$-tuples $ v $ and $ w $ of distinct elements in $ [n] $ there is a product of the $ P_i $'s of length $ O(\log n) $ such that $ P(v(j),w(j))>0 $ for all $ j=1,\dots ,r $.
\end{theorem}
We remark that Theorem \ref{thm:perm} also holds in case some of the matrices $ P_i $ are sampled (uniformly) from the set of the binary matrices that dominate at least a permutation matrix.

\begin{proof} \textbf{of Theorem \ref{Thm:primperm}.}\\
Let $ \mathcal{M}\!=\!\lbrace P_1, \bar{P}_2\rbrace $ be a random perturbed permutation set with $\bar{P}_2=P_2+ \mathbb{I}_{i,j}  $ and let $ i'\neq i $ the integer such that $ P_2(i',j)\!=\!1 $. Theorem \ref{thm:perm} with $ r=2 $ and $ d=2 $ states that property $ F_2 $ happens with high probability, i.e.\ with high probability for \emph{any} indices $ v_1,v_2,w_1,w_2\in [n] $ there exists a product $ Q $ of elements of $ \mathcal{M} $ of length $ O(\log n) $ such that $ Q(v_1,w_1)>0 $ and $ Q(v_2,w_2)>0 $.  
We now construct a product of elements of $ \mathcal{M} $ whose $ j $-th column is entrywise positive; to do so we will proceed recursively by constructing at each step a product that has one more positive entry in the $ j $-th column than in the previous step.
Matrix $ \bar{P}_2 $ has two $ 1s $ in its $ j $-th column; let $ a_1 $ and $ b_1 $ be two indices such that $ \bar{P}_2(a_1,j)=0 $ and $ \bar{P}_2(a_1,b_1)=1 $ (they do exist as the matrices are NZ). By Property $ F_2 $ there exists a product $ Q_1 $ of elements in $ \mathcal{M} $ such that $Q_1(j,i)>0$ and $ Q_1(b_1,i')>0 $; then the product $ \bar{P}_2Q_1\bar{P}_2:=K_1 $ has three positive entries in its $ j $-th column. Let now $ a_2 $ and $ b_2 $ be two indices such that $ K_1(a_2,j)\!=\!0 $ and $ K_1(a_2,b_2)>0 $; by Property $ F_2 $ there exists a product $ Q_2 $ such that $ Q_2(j,i)>0 $ and $ Q_2(b_2,i')>0 $ and so the product $ K_1Q_2\bar{P}_2:=K_2  $ has four positive entries in its $ j $-th column. 
If we iterate this procedure, it is clear that $ K_{n-2} $ has a positive column in position $ j $. Still by Property $ F_2 $, each product $ Q_i $ for $ i=1,\dots ,n-2 $ has length $ O(\log n) $, so $ K_{n-2} $ has length $ O(n\log n) $. The same reasoning can be applied to the set $ \mathcal{M}^T =\lbrace P^T_1, \bar{P}^T_2\rbrace $ since it is still a perturbed permutation matrix as $ \bar{P}^T_2=P^T_2+ \mathbb{I}_{j,i}$: there exist products $ T_1,T_2,\dots ,T_{n-2} $ of elements of $ \mathcal{M}^T $ of length $ O(\log n) $ such that, by setting $ W_1=\bar{P}^T_2T_1\bar{P}^T_2 $ and $ W_s=W_{s-1}T_s\bar{P}^T_2 $ for $ s=2,\dots ,n-2 $, the final product $ W_{n-2} $ has length $ O(n\log n) $ and its $ i $-th column is entrywise positive. Finally Property $ F_2 $ implies that there exists a product $ S $ of elements in $ \mathcal{M} $ of length $ O(\log n) $ such that $ S(j,i)>0 $. Then $K_{n-2} SW^T_{n-2} $ is a positive product of elements in $ \mathcal{M} $ of length $ O(n\log n) $.
Therefore,
\begin{equation}\label{eq:primhp}
\mathbb{P}(\mathcal{M}\text{ is primitive})\geq \mathbb{P}\bigl(\exists \text{ products } Q_1,\dots, Q_{n-1},T_1,\dots ,T_{n-2},S\text{ of length }O(\log n)\bigr)
\end{equation}
where the righ-hand side of (\ref{eq:primhp}) tends to $ 1 $ as $ n\rightarrow\infty $ by Theorem \ref{thm:perm}.
\end{proof}

\begin{proof} \textbf{of Theorem \ref{thm:binarymat}.}\\
The probability that $ B(p) $ dominates a permutation matrix is equal to the probability that a random bipartite graph admits a perfect matching; if $np-\log n\longrightarrow \infty $ this event happens with high probability (\cite{randgraphs}, Theorem 4.1). We claim that $ B(p) $ also dominates a perturbed permutation matrix with high probability if $np-\log n\longrightarrow \infty $: indeed, the probability that $ B(p) $ dominates a perturbed permutation matrix is equal to the probability that $ B(p) $ dominates a permutation matrix minus the probability that $ B(p) $ \textit{is} a permutation matrix. As $
\mathbb{P}(B(p) \text{ is a perm. matrix})\leq \left( np(1-p)^{n-1}\right) ^n$ and the right-hand side of this inequality goes to $ 0 $ as $ n $ tends to infinity by hypothesis, the claim is proven. Combining these two little results, we have proved that $ \mathcal{B}(p) $ dominates a perturbed permutation set of cardinality $ 2 $ with high probability. What remains to show is that, sampling uniformly at random a perturbed permutation set among ones that are dominated by $ \mathcal{B}(p) $ induces the same probability distribution on $ S_n\times \bar{S}_n $ as the one defined by Procedure \ref{proc}. We can then conclude by applying Theorem \ref{Thm:primperm} as, if  $ \lbrace A_2,B_2\rbrace $ is a primitive set and $ A_1,\,B_1 $ are two matrices such that $ A_1\geq A_2$ and $B_1\geq B_2 $, then $ \lbrace A_1,B_1\rbrace $ is primitive. To do so, we introduce two random variables $ X: B(p)\to S_n\cup \lbrace 0\rbrace$ and $ \bar{X}: B(p)\to  \bar{S}_n\cup \lbrace 0\rbrace  $ defined as follows: if $ B(p) $ dominates a (perturbed) permutation matrix, then  $ X $ ($ \bar{X} $) is equal to a (perturbed) permutation matrix uniformly sampled among the ones dominated by $ B(p) $, otherwise $ X $ ($ \bar{X} $) is set equal to $ 0 $. Let $ \mathbb{P}_X $ and $\mathbb{P}_{\bar{X}}  $ be the distributions of respectively $ X $ and $ \bar{X} $, 
 $  \mathbb{P}_{X\times\bar{X}}$ be the product distribution $ \mathbb{P}_X \times \mathbb{P}_{\bar{X}} $ and $ E $ the event that a set in $ S_n\times \bar{S}_n $ is primitive and with exponent of order $ O(n\log n) $; it then sufficies to prove that $ \mathbb{P}_{X\times\bar{X}}(E) $ tends to $ 1 $ as $ n $ goes to infinity. We claim that for every $ P,Q\in S_n  $ and $ \bar{P},\bar{Q}\in \bar{S}_n  $ it holds that 
\begin{equation}\label{eq:almostunif}
\mathbb{P}_X(P)=(1-\mathbb{P}_X( 0))/ n! \quad\text{ and } \quad\mathbb{P}_{\bar{X}}(\bar{P})=(1-\mathbb{P}_{\bar{X}}( 0))/(n!\,n(n-1)).
\end{equation}
Indeed, it sufficies to show that $
\mathbb{P}_X(P)=\mathbb{P}_X(Q)$ for every $ P,Q\in S_n  $ and that $\,\mathbb{P}_{\bar{X}}(\bar{P})=\mathbb{P}_{\bar{X}}(\bar{Q})$ for every $ \bar{P},\bar{Q}\in \bar{S}_n  $. We prove the first equality: 
 by definition, for any $ P\in S_n$, $\mathbb{P}_X(P)=\sum_{M\geq P}\mathbb{P}(B(p)\!=\!M) \vert \lbrace P': M\!\geq\! P'\rbrace\vert^{-1}  $, where $ \mathbb{P} $ is the distribution of $ B(p) $. Let's fix $ P,Q\in S_n  $ and let $ B $ be a realization of $ B(p) $; since $ S_n $ is transitive on itself as a group, there exist a $ T\!\in\! S_n  $ such that $ P=QT $.  
We observe that $ \mathbb{P}(B) $ depends only on the number of positive entries of $ B $ so $ \mathbb{P}(B)=\mathbb{P}(BT^{-1}) $ as $ T $ is a permutation. Then,
$
\mathbb{P}_X(P)=\sum_{B\geq QT}\,\mathbb{P}(BT^{-1})\, \vert \lbrace P': BT^{-1}\geq P'T^{-1}\rbrace\vert^{-1}=\sum_{BT^{-1}\geq Q}\,\mathbb{P}(BT^{-1})\, \vert \lbrace P': BT^{-1}\geq P'\rbrace\vert^{-1}=\mathbb{P}_X(Q).$
 The second equality can be proved by similar argument by observing that for every $ \bar{P},\bar{Q}\!\in\! \bar{S}_n  $ there exist two permutation matrices $ T_1 $ and $ T_2 $ such that $ \bar{P}\!=\! T_1\bar{Q}T_2 $. By applying equation (\ref{eq:almostunif}) it holds that:
\begin{equation}\label{eq:binmat}
\mathbb{P}_{X\times\bar{X}}(E)=\!\!\!\!\sum_{\lbrace P_1,\bar{P}_2\rbrace\in E }\!\!\!\!\mathbb{P}_{X}(P_1)\mathbb{P}_{\bar{X}}(\bar{P}_2)= \bigl(1-\mathbb{P}_{X}( 0)\bigr)\bigl(1-\mathbb{P}_{\bar{X}}( 0)\bigr)\!\!\!\sum_{\lbrace P_1,\bar{P}_2\rbrace\in E  }\!\!\! (n!)^{-1} \bigl(n!\, n(n-1)\bigr)^{-1}
\end{equation}

The summation in (\ref{eq:binmat}) is indeed the probability that a set generated by Procedure \ref{proc} is primitive and with exponent of order $ O(n\log n) $, which happens with high probability by Theorem \ref{Thm:primperm}. Finally, $ \mathbb{P}_{X}(0) $ and $ \mathbb{P}_{\bar{X}}(0) $ tend to zero  as $ n $ tends to infinity by what showed at the beginning of the proof, as $ \mathbb{P}_{X}( 0) $ ($ \mathbb{P}_{\bar{X}}(0) $) is the probability that $ B(p) $ does \emph{not} dominate a (perturbed) permutation matrix .
\end{proof}

\begin{proof} \textbf{of Proposition \ref{prop:notprim}.}\\
A set of two binary matrices $ \lbrace B_1,B_2\rbrace $ is reducible if and only if the matrix $ L=B_1\vee B_2 $ is reducible, where $ \vee $ denotes the entrywise supremum. The distribution of $ L(p) $ is the one obtained by setting in a matrix an entry equal to $ 0 $ with probability $ 1-q=(1-p)^2 $ and equal to $ 1 $ with probability $ q=p(2-p) $, independently of each others. The probability of $ L(p) $ to be reducible is equal to the probability that $ G(n,q) $ is \textit{not} strongly connected, with $ G(n,q) $ a random digraph on $ n $ vertices where a directed edge between two vertices is put, independently, with probability $ q $. Graham and Pike proved (\cite{Graham}, Corollary 1) that this happens with high probability as long as $ nq-\log n\rightarrow -\infty $. In our case, $ nq-\log n= np(2-p)-\log n\leq 2np-\log n$ which goes to $ -\infty $ by hypothesis.
\end{proof}

\end{document}